\documentclass[notoc]{JHEP3}
\usepackage{epsfig}
\usepackage{axodraw}

\def\be{\begin{equation}}  
\def\ee{\end{equation}}  

\title{Singlet Higgs Phenomenology and\\ the Electroweak Phase Transition}
\author{Stefano Profumo \\
	California Institute of Technology, Mail Code 106-38, Pasadena, CA 91125, USA, and\\
\mbox{Santa Cruz Institute for Particle Physics, University of California,
Santa Cruz, CA 95064, USA}\\
    E-mail: \email{profumo@caltech.edu}}
\author{Michael J. Ramsey-Musolf \\
	California Institute of Technology, Mail Code 106-38, Pasadena, CA 91125, USA, and\\
	University of Wisconsin-Madison, Madison, WI 53706, USA\\
	E-mail: \email{mjrm@caltech.edu}}
\author{Gabe Shaughnessy\\
	University of Wisconsin-Madison, Madison, WI 53706 USA\\
	E-mail:\email{gshau@hep.wisc.edu}}

\preprint{Caltech MAP-333\\ MADPH-07-1489}

\abstract{We study the phenomenology of gauge singlet extensions of the Standard Model scalar sector and their implications for the electroweak phase transition. We determine the conditions on the scalar potential parameters that lead to a strong first order phase transition as needed to produce the observed baryon asymmetry of the universe. We analyze the constraints on the potential parameters derived from Higgs boson searches at LEP and electroweak precision observables. For models that satisfy these constraints and that produce a strong first order phase transition, we discuss the prospective signatures in future Higgs studies at the Large Hadron Collider and a Linear Collider. We argue that such studies will provide powerful probes of phase transition dynamics in models with an extended scalar sector.}

\begin{document}

\section{Introduction}\label{sec:intro}

Explaining the origin of the baryonic matter of the universe remains an unsolved problem that lies at the interface of cosmology with particle and nuclear physics. Assuming that the universe was matter-antimatter symmetric at the end of inflation, the particle physics of the subsequently evolving universe would have to be responsible for the baryon asymmetry of the universe (BAU) observed in the present epoch: 
\be
Y_B\equiv \frac{n_B}{s_\gamma} = 
\biggl\{
\begin{array}{cl}
(7.95\pm 0.65)\times 10^{-11}, & {\rm BBN\ \cite{Yao:2006px}}\\
(9.29\pm 0.34)\times 10^{-11}, & {\rm WMAP\ \cite{Spergel:2006hy}}
\end{array}
\ee
where $n_B$ is the baryon number density, $s_\gamma$ is the photon entropy density at freeze-out,  the first value (BBN) is obtained from observed light element abundances and the predictions of standard Big Bang Nucleosynthesis, and the second value (WMAP) is extracted from the details of the acoustic peaks in the cosmic microwave background anisotropy power spectrum as probed by the WMAP collaboration. It is well-known that three ingredients must be present for successful baryogenesis \cite{Sakharov:1967dj}: the violation of total baryon number (B); violation of both C- and CP- symmetry; and a departure from thermal equilibrium\footnote{The third ingredient can be avoided if CPT is violated.}. Although the Standard Model (SM), in principle, contains all three ingredients, the strength of CP-violation is too suppressed\footnote{The strength of CP-violating effects in the electroweak sector of the SM is characterized by the Jarlskog invariant \cite{Jarlskog:1985ht}, $J=(3.08^{+0.16}_{-0.10})\times 10^{-5}$ \cite{Yao:2006px}.} and the departure from thermal equilibrium too gentle to bring about the observed value of $Y_B$ \cite{Shaposhnikov:1987tw}. 

A variety of baryogenesis scenarios involving new physics have been considered over the years, with the energy scale of the  corresponding dynamics ranging from the electroweak scale to the Planck scale (for recent reviews, see, {\em e.g.}, \cite{Riotto:1999yt,Dine:2003ax}). With the imminent start of operations at the CERN Large Hadron Collider (LHC), and the prospect for more powerful probes of new electroweak CP-violation using searches for the permanent electric dipole moments (EDM) of the electron, neutron, neutral atoms, and the deuteron \cite{Pospelov:2005pr,Erler:2004cx,Ramsey-Musolf:2006vr}, it is a particularly interesting time to scrutinize the possibilities for electroweak baryogenesis (EWB), i.e. that the BAU was generated at the electroweak phase transition (EWPT) in the early universe. The EDM searches may tell us whether or not   sufficiently large CP-violating asymmetries could have been created during the era of electroweak symmetry breaking, while LHC studies will provide information on the mechanism of EW symmetry breaking. In turn, this can lead to insights on the nature of the EWPT in the early universe, including whether, indeed, an EWPT occurred \cite{Rummukainen:1998as}.

In this paper, we consider the implications of existing and future collider studies for the extensions of the scalar sector of the SM as they bear on the EWPT. In order to prevent the \lq\lq washout" of any baryon asymmetry by electroweak sphalerons, the EWPT needs to be strongly first order -- a requirement quantitatively characterized by the condition on the scalar (Higgs) background field $\varphi=\sqrt{2}\langle H^0\rangle$
\be
\label{eq:crit1}
\frac{\varphi_c}{T_c}\gtrsim 1\ \ \ ,
\ee
where $\langle H^0\rangle$ is the vacuum expectation value (vev) of the neutral component of the SU(2)$_L$ Higgs doublet and $T_c$ is the critical temperature at which the value of the minima of the Higgs potential at $\varphi=0$ and $\varphi=\varphi_c\not=0$ become degenerate. Whether or not an EWPT satisfies Eq.~(\ref{eq:crit1}) depends in detail on the scalar sector of the electroweak theory. In the SM, Eq.~(\ref{eq:crit1}) is equivalent to the following condition on the mass of the Higgs boson, $m_h$:
\be
\label{eq:critsm}
4E_{SM}\, \left(\frac{v_0^2}{m_h^2}\right)\gtrsim 1\ \ \ ,
\ee
where $v_0\approx 246$ GeV is the the value of $\varphi$ at $T=0$, and $T E_{SM}$ is the calculable coefficient of the cubic term in the finite temperature SM scalar potential. For the value of $E_{SM}$ obtained in the SM, one finds that the Higgs mass must satisfy $m_h\lesssim 42$ GeV, in conflict with the direct search lower bound of 114.4 GeV from LEP 2~\cite{Barate:2003sz}.

Extensions of the scalar sector of the SM can relax this requirement on $m_h$. It is well known, for example, that in the Minimal Supersymmetric Standard Model (MSSM), the presence of a light stop can enhance the cubic term, leading to $E_{MSSM}\approx10\,  E_{SM}$, allowing the mass of the lightest, SM-like Higgs to be as heavy as $\sim 120$ GeV \cite{ewbstop}, and thereby compatible with direct collider searches \cite{MSSMHiggs}. This enhancement arises when the soft mass $M_U$ is chosen to nearly cancel the finite temperature mass function $\Pi_{\tilde t}(T)$. In this case,  the order of magnitude increase in $E$ arises largely from the number of stop degrees of freedom ($=2N_C$). Alternately, numerical studies have shown that extensions of the SM scalar sector containing singlet fields can also strengthen the EWPT \cite{Pietroni:1992in,othersinglet,langauprime} so that Eq.~(\ref{eq:crit1}) is satisfied with scalar masses above the current direct search lower bound for the Higgs \cite{Barate:2003sz}. Such singlets may be the low-energy remnants of scalar multiplets that are charged under gauge symmetries  broken at high scales \cite{langauprime}. In supersymmetric models, they provide an attractive solution to the $\mu$-problem, and models with dynamically generated $\mu$-terms have been studied extensively in the literature \cite{muproblem}. A singlet field coupled to right-handed neutrinos can also give rise to a non-thermal production mechanism for sterile neutrino dark matter accommodating an explanation of both the cosmological dark matter and the observed velocities of pulsars \cite{singletkusenko}.

In this paper, we analyze the generic implications of singlet extensions of the SM scalar sector for the EWPT and discuss the corresponding implications for low-energy phenomenology. Our goal is to identify  generic features of singlet extensions that are relevant to the EWPT and to low-energy Higgs phenomenology, and that might arise   in specific realizations, such as the above mentioned singlet extensions of the MSSM.
To that end, we consider the minimal extension of the SM scalar sector containing one additional real, singlet scalar field, $S$ \cite{O'Connell:2006wi}. Although this scenario does not contain additional CP-violation beyond that of the SM and, thus, does not directly address the new CP-violation needed for successful EWB, it does provide a framework for understanding general characteristics of the EWPT in models with extra singlet scalars. We will assume that the need for new CP-violation is addressed by aspects of a given SM extension that may or may not be directly related to the augmented scalar sector. 

In performing our study, we rely on a combination of analytic and numerical analysis. In doing so, we obtain an analog to Eq.~(\ref{eq:critsm}) that applies for phenomenologically viable parameters:
\be
\label{eq:crit2}
\sqrt{2}\cos\alpha_c\, \left(\frac{\varepsilon-e/T_c}{2\bar\lambda}\right)\, \left[1+\gamma\frac{|V_0|}{T_c^4}\right] +\cdots \gtrsim 1\ \ \ ,
\ee
where $\tan\alpha_c$ is the ratio of the singlet vev and $v/\sqrt{2}$ at $T_c$; $T\varepsilon$  and $e$ are coefficients, respectively,  of the $T$-dependent and $T$-independent cubic terms in the finite temperature effective potential; ${\bar\lambda}$ is the coefficient of the effective quartic term; $V_0$ is the value of the degenerate minima of the potential at $T_c$, defined with respect to the value at the origin;  $\gamma$ is a computable, positive constant that depends on the parameters in the potential; and the $+\cdots$ denote non-analytic or subdominant finite temperature contributions. 

From Eq.~(\ref{eq:crit2}), one sees that in the present context the strength of the EWPT can be enhanced with respect to the SM case in one of five ways: 
\begin{itemize}
\item[(a)] increasing $\varepsilon$ compared to its SM value ($\varepsilon_{SM}=2\sqrt{2}E_{SM}$)
\item[(b)] introducing new cubic terms in the potential giving a negative value for $e$
\item[(c)] reducing the denominator through a  tree level effects arising from quartic interactions of the singlet and SM scalar fields 
\item[(d)] reducing the denominator through singlet contributions to one-loop renormalization of the SU(2)$_L$ quartic coupling
\item[(e)] generating a non-zero value for $V_0$ just above the critical temperature. 
\end{itemize}
The first option generally requires introduction of a large number of additional scalar degrees of freedom in order to be compatible with $m_h> 114.4$ GeV, since the finite temperature cubic term in the potential arises at loop level and is $1/4\pi$ suppressed. Similarly, option (d) requires either large singlet-doublet quartic interactions or multiple singlet scalars to overcome the one-loop suppression factors. Indeed, in the case we consider here of a single extra singlet degree of freedom, the singlet enhancements of $\varepsilon$ generally do not yield an appreciable strengthening of the EWPT (unless a large number of them are introduced, see Ref.~\cite{Espinosa:2007qk} for a recent discussion of this case). In contrast, mechanisms (b) and (c) can yield a strong first order EWPT and phenomenologically viable scalar masses with inclusion of only one new singlet scalar having  perturbative couplings to the doublet scalars, as they rely primarily on the presence of new tree-level terms in the potential. We also find numerous models that are compatible with low-energy phenomenology and that enhance the strength of the EWPT through mechanism (e).

In addition to outlining the derivation of Eq.~(\ref{eq:crit2}), we also discuss the low-energy probes of these mechanisms for enhancing the EWPT. In particular, we show that mechanism (b) can be constrained by studies of Higgsstrahlung in $e^+e^-$ annihilation, since the tree-level cubic term introduces mixing between the SM and singlet scalars. We find that Higgs boson searches at LEP already impose stringent constraints on this mechanism, but for scalar masses lying above the LEP lower bound, there exists considerable room for inducing a strong first order EWPT in this way. Existing data from electroweak precision observables (EWPO) lead to additional constraints. It is well known that global analyses of EWPO favor a rather light 
scalar\footnote{However, the forward-backward asymmetry for $e^+e^-\to Z^0\to b{\bar b}$ favors a heavy Higgs scalar, while the left-right asymmetries in $e^+e^-\to Z^0\to\, {\rm hadrons}$ indicate a Higgs with mass well below the LEP direct search bound \cite{wjm06}.}, 
and we find that models having a heavy singlet scalar that mixes strongly with the SU(2)-like scalar are disfavored. More generally, EWPO restrict the mass of the SU(2)-like scalar to values below $\sim$200 GeV while favoring small singlet masses and large mixing between light singlet-like and the SU(2)-like scalars. 

Future studies of Higgs boson production and decays at the LHC and a Linear Collider will provide powerful probes of scalar singlet models that generate a strong first order EWPT. For models in which the decay of one neutral scalar mass eigenstate into a pair of the second is kinematically allowed, one would expect deviations from SM Higgs decay branching ratios at the LHC. This possibility is particularly interesting in the case of mechanism (c) that is present in models having a $\mathbb{Z}_2$  ($S\leftrightarrow -S$, $H\leftrightarrow -H$) symmetry before spontaneous symmetry breaking (SSB).  We find that the decay of the SM-like scalar into two singlet-like scalars is always kinematically allowed for models giving rise to a strong first order EWPT, and that the branching ratio to SM final states can be significantly reduced (from unity) for relatively light scalar masses. Thus, if a SM-like Higgs is discovered at the LHC, studies of its decay may yield important constraints on the viability of EWB in models with a $\mathbb{Z}_2$ symmetry. 

For models that do not have this symmetry, it is likely that either the singlet-like scalar can decay into two SU(2)-like scalars, or that neither scalar can decay into a pair of the other. One could probe the former class by searching for exotic final states at the LHC, such as four $b$-jets or $b{\bar b}\gamma\gamma$.  For models wherein no new scalar decay channels arise, future Higgsstrahlung studies at a 500 GeV $e^+e^-$ Linear Collider could provide an additional window on EWPT-viable scenarios. We discuss both possibilities below.

Our analysis of the EWPT that leads to these phenomenological implications builds on extensive, previous work by others, both in the case of singlet scalar models \cite{Espinosa:2007qk,Anderson:1991zb,extrasinglet} and supersymmetric models with an extended scalar sector \cite{Pietroni:1992in,othersinglet}. In carrying out the present study, we have tried to amplify on earlier work by identifying general features of singlet scalar extensions that lead to a strong first order EWPT and that should apply to a broad array of specific models. We have also attempted to identify the ways in which electroweak precision measurements as well as present and future Higgs boson studies could be used to test EWPT-viable singlet extensions of the Standard Model scalar sector. Finally, we carry out a more detailed study of the pattern of symmetry-breaking associated with the singlet scalar, since models in which $\langle S\rangle\not=0$ occur quite copiously.

The discussion of our analysis is organized as follows: In Section \ref{sec:ssm} we outline the model for the singlet scalar extension of the SM, largely following the notation of Ref.~\cite{O'Connell:2006wi}. In Section \ref{sec:highT} we analyze the pattern of symmetry breaking at temperatures for which SU(2)$_L$ is unbroken but the singlet scalar obtains a non-zero vev. Section \ref{sec:ewpt} gives our analysis of electroweak symmetry breaking and the implications for the strength of the EWPT. In Section \ref{sec:pheno} we analyze the implications of low-energy (zero temperature) phenomenology for the EWPT in this scenario, including present and prospective constraints from $e^+e^-$ annihilation, electroweak precision observables (EWPO), and Higgs production and decay at the LHC. We summarize our main conclusions in Section \ref{sec:conclusions}

\section{Singlet Scalar Extension of the SM}
\label{sec:ssm}

Following Ref.~\cite{O'Connell:2006wi} (and references therein), we study a minimal extension of the Standard Model (SM) scalar sector encompassing a single gauge singlet real scalar field $S$. The associated zero temperature, tree-level scalar potential reads
\begin{equation}
\label{eq:vtree}
V=V_{\rm SM}+V_{\rm HS}+V_{\rm S}
\end{equation}
where
\begin{eqnarray}
\label{eq:potential}
&&V_{\rm SM}=-\mu^2\left(H^\dagger H\right)+\bar\lambda_0\left(H^\dagger H\right)^2\\
\nonumber &&V_{\rm HS}=\frac{a_1}{2}\left(H^\dagger H\right)S+\frac{a_2}{2}\left(H^\dagger H\right)S^2\\
\nonumber &&V_{\rm S}=\frac{b_2}{2}S^2+\frac{b_3}{3}S^3+\frac{b_4}{4}S^4\ \ \ ,
\end{eqnarray}
where $H$ is the standard SU(2)$_L$ scalar doublet with hypercharge one. Our notation follows a practice sometimes used in the literature wherein
no distinction is made between dimensionless and dimensionfull couplings
(see, {\em e.g.}, Ref. \cite{O'Connell:2006wi}). Specifically, we use the symbol $a$ for mixed singlet-SU(2)$_L$ terms and $b$ for singlet terms, and the subscripts refer to the corresponding power of the singlet field $S$. Tab.~\ref{tab:coeff} should help the reader keep track of the various coefficients appearing in $V_{\rm HS}$ and in $V_{\rm S}$, of their mass dimension and whether or not they are $\mathbb{Z}_2$ symmetric.

\begin{table}[!b]
\begin{center}
\begin{tabular}{|c|c|c|c|}\hline
Coefficient & Corresp. Term & Mass Dimension & $\mathbb{Z}_2$ symmetric\\
\hline
$a_1$ & $\left(H^\dagger H\right)S/2$ & 1 & No \\
$a_2$ & $\left(H^\dagger H\right)S^2/2$ & 0 & Yes \\
$b_2$ & $S^2/2$ & 2 & Yes \\
$b_3$ & $S^3/3$ & 1 & No \\
$b_4$ & $S^4/4$ & 0 & Yes \\
\hline
\end{tabular}
\end{center}
\caption{A summary of the coefficients employed in the potential under consideration} \label{tab:coeff}
\end{table}

We do not include a linear term in the singlet potential $V_{\rm S}$, since for a generic potential
\begin{equation}
\tilde V_{\rm S}=\tilde b_1 S+\frac{\tilde b_2}{2}S^2+\frac{\tilde b_3}{3}S^3+\frac{\tilde b_4}{4}S^4
\end{equation}
one can always remove the linear term by shifting $S\rightarrow S-\beta$, where $\beta$ is a real solution to the equation
\begin{equation}
\label{eq:vshift}
\tilde b_1-\tilde b_2\beta+\tilde b_3\beta^2-\tilde b_4\beta^3=0
\end{equation}
and obtain the  $b_i$ coefficients in Eq.~(\ref{eq:potential}) from
\begin{eqnarray}
\nonumber
&&b_2\equiv \tilde b_2-2\tilde b_3\beta+3\tilde b_4\beta^2\\
\label{eq:bshift}
\nonumber &&b_3\equiv \tilde b_3-3\tilde b_4\beta\\
\nonumber
\nonumber &&b_4\equiv \tilde b_4\ \ \ .
\end{eqnarray}

In contrast to Ref.~\cite{O'Connell:2006wi}, we have not shifted $S$ to remove its vev because we are interested in the pattern of symmetry breaking in the model at non-vanishing temperatures. Thus, we will keep the dependence on the neutral scalar vevs explicit. Letting $v_0/\sqrt{2}$ and $x_0$ be the $T=0$ vacuum expectation values (vevs) of the neutral component of the SM Higgs $H^0$ and singlet scalar $S$, respectively,
the potential minimization conditions
\begin{eqnarray}
\frac{\partial V}{\partial H}\Bigg|_{\langle H^0\rangle_{T=0}=v_0/\sqrt{2}}\ =\ \frac{\partial V}{\partial S}\Bigg|_{\langle S\rangle_{T=0}=x_0}=0
\end{eqnarray}
allow us to express two of the parameters appearing in Eq. (\ref{eq:potential}) in terms of the zero temperature vevs and the other parameters in the potential. For $x_0\neq 0$, we choose to eliminate the mass parameters $\mu^2$ and $b_2$, 
\begin{eqnarray}
\label{eq:mu2}
&&\mu^2=\bar\lambda_0 v^2_0+\left(a_1+a_2x_0\right)\frac{x_0}{2}\\
\label{eq:b2}
&&b_2=-b_3x_0-b_4x^2_0-\frac{a_1 v^2_0}{4x_0}-\frac{a_2 v^2_0}{2}.
\end{eqnarray}
For $x_0=0$, the minimization condition with respect to $S$ enforces the condition $a_1=0$, while $\mu^2=\bar\lambda_0 v^2_0$ as in the SM. In this case, as we discuss below, this is the necessary and sufficient condition for having a stable neutral scalar that can be the DM, as first noticed Ref.~\cite{Silveira:1985rk}. We emphasize, however, that imposing a tree-level $\mathbb{Z}_2$ symmetry on the potential ($a_1=0=b_3$) does not imply a vanishing singlet vev. Only when $x_0=0$ is it possible to have a stable neutral scalar. While this assumption is implicit in many previous analyses, we find that models with $x_0\neq 0$ arise copiously in the present framework. 

The fields  ($h$, $s$) describing fluctuations about the vevs are defined by  $H^0=(v_0+h)/\sqrt{2}$ and $S=x_0+s$, at $T=0$. The corresponding entries in the mass matrix are given by\footnote{We discuss corrections resulting from the full Coleman-Weinberg effective potential below. These corrections lead to numerically small shifts to these conditions.}
\begin{eqnarray}
\mu^2_h\equiv\frac{\partial^2 V}{\partial h^2}&=&2\bar\lambda_0 v^2_0\\
\mu^2_s\equiv\frac{\partial^2 V}{\partial s^2}&=&b_3x_0+2b_4x_0^2-\frac{a_1v^2_0}{4x_0}\\
\label{eq:massparams}
\mu^2_{hs}\equiv\frac{\partial^2 V}{\partial h\partial s}&=&\left(a_1+2a_2x_0\right)\, v_0\ \ \ .
\end{eqnarray}
The mass eigenstates $h_1$ and $h_2$ are defined as
\begin{eqnarray}
\nonumber
&&h_1=\sin\theta\ s+\cos\theta\ h\\
\label{eq:eigenstates}
&&h_2=\cos\theta\ s-\sin\theta\ h
\end{eqnarray}
where the mixing angle $\theta$ is given by
\begin{equation}
\label{eq:mixing}
\tan\theta=\frac{y}{1+\sqrt{1+y^2}},\qquad{\rm where}\quad y\equiv\frac{\mu^2_{hs}}{\mu^2_h-\mu^2_s}.
\end{equation}
With this convention, $|\cos\theta|>1/\sqrt{2}$, therefore $h_1$ is the
mass eigenstate with the largest SU(2)-like component and $h_2$ that with
the largest singlet component. The corresponding mass eigenvalues are given by
\begin{eqnarray}
m^2_{1,2}=\frac{\mu^2_h+\mu^2_s}{2}\pm\frac{\mu^2_h-\mu^2_s}{2}\sqrt{1+y^2} \ \ \
\end{eqnarray} 
where the upper (lower) sign corresponds to $m_1$ ($m_2$). 

For future reference it is useful to relate the parameters in $V$ to those appearing in the notation of Ref.~\cite{O'Connell:2006wi}, where the potential is written in terms of the zero-temperature, shifted field $s$ only. One has
\begin{eqnarray}
\label{eq:shiftpot}
V(H,s) & = & -\frac{\mu^2_h}{2}\left(H^\dagger H\right)+\bar\lambda_0\left(H^\dagger H\right)^2+\frac{\delta_1}{2}\left(H^\dagger H\right) s \\
\nonumber
& & + \frac{\delta_2}{2} (H^\dag H)s^2-\left(\frac{\delta_1\mu^2_h}{{8\bar\lambda_0}}\right)s+\frac{\kappa_2}{2}s^2+\frac{\kappa_3}{3} s^3 +\frac{\kappa_4}{4} s^4\ \ \ ,
\end{eqnarray}
where 
\begin{eqnarray}
\nonumber
\delta_1 & = & a_1+2a_2 x_0\\
\nonumber
\delta_2 & = & a_2 \\
\label{eq:shiftparams}
\kappa_2 & = &  b_2+2b_3 x_0+3 b_4 x_0^2 \\
\nonumber
& = & b_3 x_0 +2b_4 x_0^2-\frac{a_1 v^2}{4 x_0}-\frac{a_2v^2}{2}\\
\nonumber
\kappa_3 & = & b_3 + 3 b_4 x_0 \\
\nonumber
\kappa_4 & = & b_4 \ \ \ ,
\end{eqnarray}
and where we have used the condition for a non-zero $x_0$ to eliminate $b_2$ in terms of $x_0$ and the other parameters.
From Eqs.~(\ref{eq:massparams}-\ref{eq:shiftparams}) we observe that even if the potential displays a $\mathbb{Z}_2$  ($S\leftrightarrow -S$, $H\leftrightarrow -H$) symmetry before spontaneous symmetry-breaking, the zero temperature potential will not in general do so if the singlet vev $x_0\not=0$. In this case, one would encounter mixing between the neutral SU(2$)_L$ and singlet scalars, thereby allowing for decay of both mass eigenstates $h_1$ and $h_2$ to SM particles and precluding either from being a viable dark matter candidate. Only if  $a_1$, $x_0$, and $b_3$ all  vanish does one allow for a (light) stable neutral scalar\footnote{For non-vanishing $b_3$, the singlet scalar can decay to a pair of SU(2)$_L$ scalars at one-loop order.}. 

Various theoretical and phenomenological criteria restrict the values of the parameters in $V$:
\begin{itemize}
\item[(i)] Theoretically, $V$ must be bounded from below. This condition requires the positivity of the quartic coefficients $b_4$ and $\bar\lambda_0$. In addition, requiring the positivity of the quartic coefficient along an arbitrary direction implies that, if $a_2<0$, $a_2^2<4b_4\bar\lambda_0$. Alternatively, one can require positivity of the determinant of the bilinear form including terms containing four powers of the fields, obtaining the same condition.
\item[(ii)] Electroweak symmetry breaking is viable if (1) the determinant of the mass matrix is positive and (2) the electroweak vacuum is the absolute minimum of the potential. The first (necessary, but not sufficient) condition can be cast as
\begin{equation}
b_3x+2b_4x^2-\frac{a_1v^2}{4x}-\frac{\left(a_1+2a_2x\right)^2}{8\bar\lambda_0}>0\ \ \ ,
\end{equation}
where $v$ and $x$ denote the vevs at arbitrary $T\le T_c$, with $T_c$ being the temperature at which the EWPT takes place. The second condition  (that the electroweak minimum is the absolute minimum of the potential at $T\le T_c$) cannot be cast in a simple analytic form:  at a given temperature, the potential $V$ in general has various minima. In our analysis below, we impose this condition numerically computing $V$ at all minima and requiring that the electroweak vacuum is the true vacuum of the theory.
\item[(iii)] Phenomenologically, searches for the Higgs boson constrain the combinations of parameters that determine the masses $m_{1,2}$ and $\tan\theta$. The LEP Higgs search results allow the existence of a scalar with mass below 114.4 GeV if its coupling to the $Z$-boson $g_{HZZ}$ is reduced from the SM coupling, $g_{HZZ}^{\rm SM}$. The corresponding Higgsstrahlung rate will be reduced by $\xi^2=(g_{HZZ}/g_{HZZ}^{\rm SM})^2$ compared to the SM rate if the decay branching ratios the same as in the SM. For the present case, this reduction factor is given by $\xi^2_1=\cos^2\theta$ and $\xi^2_2=\sin^2\theta$ for the two mass eigenstates $h_1$ and $h_{2}$, respectively. If the decay $h_1\to h_2 h_2$ or $h_2\to h_1 h_1$ becomes kinematically allowed, then one has
\be\label{eq:xi}
\xi^2_i=\left(\frac{g_{HZZ}}{g_{HZZ}^{\rm SM}}\right)^2 \, \frac{\Gamma^{\rm SM}}{\Gamma^{\rm SM}+\Gamma(h_i\to h_j h_j)}\ \ \ ,
\ee
where $\Gamma^{\rm SM}$ is the total decay width of of $h_i$ into conventional SM Higgs decay channels at the relevant mass.
In our numerical scans discussed below, we will impose the LEP constraints in the $(m_{k}, \xi^2_{k})$ plane ($k=1,2$), as given in Ref.~\cite{Barate:2003sz}.
\end{itemize}

\subsection{The finite temperature effective potential}

For purposes of analyzing the pattern of symmetry breaking involving both $H^0$ and $S$, it is convenient to follow Pietroni \cite{Pietroni:1992in} and work with a cylindrical coordinate representation of the classical fields $v(T)$ and $x(T)$. To that end, we define $\varphi\equiv\varphi(T)$ and $\alpha\equiv\alpha(T)$ via
\begin{eqnarray}
v/\sqrt{2}&\equiv &\varphi\cos\alpha\\
\nonumber x&\equiv&\varphi\sin\alpha \ \ \ .
\end{eqnarray}
The resulting tree-level effective potential is 
\begin{eqnarray}
\label{eq:potential2}
V_{0}^{T=0}(\varphi,\alpha)&=&\left(\mu^2\cos^2\alpha-\frac{b_2}{2}\sin^2\alpha\right)\varphi^2\\
\nonumber &+&\left(\frac{a_1}{2}\cos^2\alpha+\frac{b_3}{3}\sin^2\alpha\right)\sin\alpha\, \varphi^3\\
\nonumber &+&\left(\bar\lambda_0\cos^4\alpha+\frac{a_2}{2}\cos^2\alpha\sin^2\alpha+\frac{b_4}{4}\sin^4\alpha\right)\varphi^4\ \ \ ,
\end{eqnarray}
where $\mu^2$ and $b_2$ are implicit functions of the $T=0$ scalar vevs and of the other parameters in the potential as per Eqs.~(\ref{eq:mu2},\ref{eq:b2}).

The  $T=0$ Coleman-Weinberg one-loop effective potential can be expressed in terms of the field-dependent masses $m_i(v,x)\equiv m_i(\varphi, \alpha)$ \cite{Coleman:1973jx}:
\begin{eqnarray}
V_1^{T=0} & = & \sum_k n_k\, G\left[m_k^2(\varphi, \alpha)\right]\\
\nonumber
G(y) & = & \frac{y^2}{64\pi^2}\left[\ln\left(\frac{y}{Q^2}\right)-\frac{3}{2}\right]\ \ \ ,
\end{eqnarray}
where $Q$ is a renormalization scale (which we fix to $Q=v_0$), the sum is over all fields that interact with the fields $h$ and $S$, and $n_k$ is the number of degrees of freedom for the $k$-th particle, with a minus sign for fermionic particles. The inclusion of $V_1^{T=0}$ leads to a shift in the minimization conditions\footnote{The correction to the mass parameters also occurs, as noted above.}, corresponding to a shift in the dependence of $\mu^2$ and $b_2$ on the zero temperature vevs and other parameters: 
\begin{eqnarray}
\Delta\mu^2 & = & \frac{1}{v}\, \frac{\partial}{\partial v}\, V_1^{T=0} \\
\nonumber \Delta b_2 & = & -\frac{1}{x}\, \frac{\partial}{\partial x}\, V_1^{T=0}\ \ \ .
\end{eqnarray}
Note also that the term proportional to $-3/2y^2$ in $G(y)$ leads to a reduction in the magnitude of the quartic couplings $\bar\lambda_0$ and $b_4$ from their tree level values~\cite{Anderson:1991zb}.

The finite temperature component of the effective potential receives two contributions: the one-loop component, $V_1^{T\not=0}$ and the bosonic ring contribution $V_{\rm ring}$ that depends on the boson thermal masses. For the simple model we consider here, both contributions have been given in Refs.~\cite{Dolan:1973qd,Ahriche:2007jp} and we include those results here for completeness:
\be
V_1^{T\not=0}  =  \frac{T^4}{2\pi^2}\, \sum_k n_k J_{B,F}\left[m_k^2(\varphi,\alpha)/T^2\right]
\ee
where the functions $J_{B,F}(y)$ are given by 
\be
J_{B,F}(y) = \int_0^\infty\, dx\, x^2\, \ln\left[1\mp {\rm exp}\left(-\sqrt{x^2+y}\right)\right]
\ee
withe the upper (lower) sign corresponding to bosonic (fermionic) contributions. Similarly, one has 
\be
V_{\rm ring}(\varphi,\alpha, T) = -\frac{T}{12\pi}\, \sum_k\, n_k\, \left\{ \left[M_k^2(\varphi,\alpha,T)\right]^{3/2}-\left[m_k^2(\varphi,\alpha)\right]^{3/2}\right\}\ \ \ ,
\ee
where the thermal masses $M_k^2(\varphi,\alpha,T)$ are given in terms of the $T=0$ masses and the finite temperature mass functions, $\Pi_k$:
\be
M_k^2(\varphi,\alpha,T)=m_k^2(\varphi,\alpha)+\Pi_k\ \ \ .
\ee
Explicit expressions for the $\Pi_k$ are given in Ref.~\cite{Ahriche:2007jp}, and we do not reproduce them explicitly here. 

When $T$ is large compared to a given mass $m_k$, it is convenient to expand the functions $J_{B,F}$ \cite{Dolan:1973qd}
\begin{eqnarray}
J_{B}(y) & \approx & -\frac{\pi^4}{45}+\frac{\pi^2}{12}\, y^2 -\frac{\pi}{6}\, y^3 -\frac{y^4}{32}\ln\left(\frac{y^2}{a_B}\right)\\
\nonumber J_{F}(y) & \approx & \frac{7 \pi^4}{360}-\frac{\pi^2}{24}\, y^2 -\frac{y^4}{32}\ln\left(\frac{y^2}{a_F}\right)\ \ \ ,
\end{eqnarray}
where $a_B=16\pi^2{\rm exp}(3/2-2\gamma_E)$, $a_F=\pi^2{\rm exp}(3/2-2\gamma_E)$, and $\gamma_E$ is the Euler constant. The resulting expression for the full effective potential is given by
\be
\label{eq:vTeff}
V_{\rm eff}(\varphi,\alpha,T) = V_0(\varphi,\alpha)+V_1^{T=0}(\varphi,\alpha)+\Delta V(\varphi,\alpha,T)
\ee
with
\begin{eqnarray}
\Delta V(\varphi,\alpha,T)&=&V_1^{T\not=0}(\varphi,\alpha,T)+V_{\rm ring}(\varphi,\alpha, T)\\
\nonumber &\approx & \left(-\frac{\pi^2}{90}N_B+\frac{7\pi^2}{720}N_F\right) T^4\\
\nonumber &+& \frac{T^2}{24}\sum_{k=B,F} g_k m^2_k(\varphi,\alpha)-\frac{T}{12\pi}\sum_{k=B} n_k M_k^3(\varphi,\alpha,T)\\
\nonumber &-&\frac{1}{64\pi^2}\sum_{k=B,F}\, n_km_k^4(\varphi,\alpha)\ln\left[\frac{m_k^2(\varphi,\alpha)}{a_kT^2}\right]\ \ \ ,
\end{eqnarray}
where $g_k=n_k$ for bosonic degrees of freedom and $g_k=n_k/2$ for fermionic degrees of freedom and where $N_B$ ($N_F$) denote the total number of bosonic (fermionic) degrees of freedom, where the sum extends to the three Goldstone bosons as well. 

In the singlet extension of the SM of interest here, as in the SM, the transverse components of the $W$ and $Z$ bosons receive no finite temperature corrections to the masses, so that 
\begin{eqnarray}
\left[M_W^2(\varphi,\alpha, T)\right]^T &=& \frac{g^2}{4} v^2 = \frac{g^2}{2}\, \cos^2\alpha\, \varphi^2\\
\nonumber \left[M_Z^2(\varphi,\alpha, T)\right]^T &=& \frac{g^2+g^{\prime\ 2}}{4} v^2 = \frac{g^2+g^{\prime\ 2}}{2}\, \cos^2\alpha\, \varphi^2\ \ \ .
\end{eqnarray}
In contrast, all other particle masses are screened by non-zero $\Pi$-functions that are independent of the background fields and proportional to $T^2$. Thus, the transverse components of the $W$ and $Z$ yield the dominant contributions to the effective cubic term in the $V_{\rm eff}(\varphi,\alpha,T)$ proportional to $T\varphi^3$. In principle, it is possible  to mitigate the screening of the scalar masses by a suitable choice of the tree-level parameters, thereby yielding additional contributions to the $T\varphi^3$ term. This strategy is similar to the case of the MSSM with a light stop, wherein the tree-level soft mass is chosen to cancel the thermal contribution.  In the present case, however, we find that following a similar strategy for the thermal scalar masses generally leads to a potential unbounded from below, nonperturbative couplings in the potential, or special, finely-tuned choices for the input parameters (such as $x_0$). In a similar vein, Espinosa and Quiros (see Ref.~\cite{extrasinglet}), studied a related class of models containing an additional complex scalar field, $\phi$, having no vev and a potential that is invariant under $\phi\to e^{i\alpha}\phi$. These authors observed that it is possible to choose the parameters in order to overcome plasma screening effects and enhance the strength of the EWPT. Nevertheless, they found that the maximum mass of the neutral SU(2)$_L$ scalar consistent with a strong first order EWPT is $\approx 80$ GeV when the dimensionless couplings in the potential are perturbative. Consequently, we will concentrate on the implications of terms in the $T=0$ potential that enhance the strength of the EWPT.

We take here as the six free parameters the potential parameters $a_1,\ a_2,\ b_3,\ b_4$, the quartic SM Higgs coupling $\bar\lambda_0$ and the singlet vev at $T=0$, $x_0$. In the numerical scans we carry out in the present analysis, we linearly sample the above parameters in the ranges given in Tab.~\ref{tab:param}. The ranges on the dimensionless parameters $\lambda$, $b_4$, and $a_2$ are
roughly consistent with the requirements of perturbativity as discussed in
Refs.~\cite{extrasinglet,espinosa,Quiros:1999jp}. In the case of the
Standard Model, these considerations imply that ${\bar\lambda_0}/g^2
\lesssim 1$, where $g^2\approx 0.4$. One may obtain further restrictions on
the value of $a_2$ by requiring perturbativity up to a scale $\Lambda$ that
defines the limit of the validity of the theory, as has been implemented by
Espinosa and Quiros in the case of the complex scalar model  mentioned above
\cite{extrasinglet}. A detailed analysis of perturbativity considerations
goes beyond the scope of the present study, where our focus falls on
identifying general trends and for which limiting the ranges of the
dimensionless parameters as in Table 1 should suffice\footnote{As we also
discuss below, imposing constraints from EWPO restricts the scalar masses to
be rather light, corresponding to relatively small magnitudes for the
dimensionless couplings in the potential. Choosing these couplings near the
upper ends of the ranges in Table 1 generally leads to scalar masses that
are inconsistent with EWPO.}.

\begin{table}
\begin{center}
\begin{tabular}{|c|c|c|c|c|c|}\hline
$a_1/{\rm GeV}$ & $a_2$ & $b_3/{\rm GeV}$ & $b_4$ & $x_0/{\rm GeV}$ & $\lambda$\\
\hline
 $[-1000;1000]$ & $[-1;1]$ & $[-1000;1000]$ & $[0;1]$ & $[0;1000]$ & $[0;1]$ \\
\hline
\end{tabular}
\end{center}
\caption{Ranges for the parameters used in our scans to generate the models employed for the figures in the remainder of the paper.} \label{tab:param}
\end{table}

\section{The Singlet vacuum before the EWPT}
\label{sec:highT}

Recent studies of the pattern of EWPT in models with an extended scalar sector include both analyses where the singlet vev was assumed to be zero \cite{Anderson:1991zb,Espinosa:2007qk} and where the possibility of a non-singlet singlet vev prior to EWSB was studied numerically \cite{Ahriche:2007jp,Ham:2004cf}. In the present section, we address -- both analytically and numerically -- the role of a non-vanishing singlet vev on the strength of the EWPT.

Spontaneous symmetry breaking in the singlet sector offers two possible scenarios: (i) the singlet scalar  has already acquired a vev before the EWPT (i.e. at $T>T_c$), or (ii) the singlet has zero vev. We show below that under reasonably general assumptions, it is possible to derive an analytical condition on the parameters appearing in the potential to determine which of the two options occurs. This is necessary in order to determine the value of the EWPT critical temperature, $T_c$, that is defined as the temperature at which the minima of broken and unbroken electroweak symmetry become degenerate. It is possible that the minimum of unbroken electroweak symmetry at $T\gtrsim T_c$ 
corresponds to a non-vanishing singlet vev, so it is useful to determine the conditions under which this possibility occurs as well as the value of the corresponding unbroken minimum. 

To that end, we consider the effective potential for $\alpha=\pi/2$ for which the SU(2)$_L$ vev is zero. It has the general form
\be
\label{eq:vsinglet1}
V_{\rm eff}(\varphi,\pi/2,T)={\bar b}_0+ {\bar b}_1\, \varphi +\frac{{\bar b}_2}{2}\, \varphi^2 +\frac{{\bar b}_3}{3}\, \varphi^3+\frac{{\bar b}_4}{4}\, \varphi^4 +\cdots\ \ \ ,
\ee
where the ${\bar b}_i$ are $T$-dependent functions of the tree-level parameters
\begin{eqnarray}
{\bar b}_0 & = & c_0\, T^4+d_0\, T^2 \\
{\bar b}_1 & = & \frac{T^2}{48}\Biggl\{ 3\left(1-\frac{3\sqrt{\alpha_\chi}}{\pi}\right)\, a_1
+\left(1-\frac{3\sqrt{\alpha_+}}{\pi}\right)\, a_1\\
&&\nonumber +4\left(1-\frac{3\sqrt{\alpha_-}}{\pi}\right)\, b_3\Biggr\}\\
{\bar b}_2 & = & b_2+\frac{T^2}{24}\Biggl\{ 3\left(1-\frac{3\sqrt{\alpha_\chi}}{\pi}\right)\, a_2
+\left(1-\frac{3\sqrt{\alpha_+}}{\pi}\right)\, a_2\\
&&\nonumber +6\left(1-\frac{3\sqrt{\alpha_-}}{\pi}\right)\, b_4\Biggr\}\\
{\bar b}_3 & = & b_3\\
{\bar b}_4 & = & b_4\ \ \ ;
\end{eqnarray}
where the \lq\lq $+\cdots$" denote non-analytic contributions arising from the logarithmic field dependence in $V_1$; where the dependence of $c_0$ and $d_0$ on the parameters in the potential is not important for this discussion and, therefore, not shown explicitly; and where we have not explicitly included the analytic $\varphi^2$ and $\varphi^4$ contributions arising from $V_1^{T=0}$ that are suppressed by $1/64\pi^2$. The quantities $\sqrt{\alpha_i}$ have been obtained by expanding the thermal masses $M_k(\varphi, \pi/2,T)$ in powers of $m_k(\varphi,\pi/2)/T$ to second order and are given by 
\begin{eqnarray}
&&\alpha_\chi=\frac{g^2}{4}+\frac{\bar\lambda_0}{2}+\frac{y_t^2}{4}+\frac{a_2}{12}\\
\nonumber &&\alpha_{+}=\frac{g^2}{4}+\frac{3\bar\lambda_0}{2}+\frac{y_t^2}{4}+\frac{a_2}{12}\\
\nonumber &&\alpha_{-}=\frac{b_4}{4}+\frac{a_2}{3}
\end{eqnarray}


The potential in (\ref{eq:vsinglet1}) can in principle develop a global minimum at $\varphi\not=0$. In order to analyze the latter possibility, we first make the assumptions that (a) the non-analytic $\varphi$-dependence arising from the screened masses and logarithmic term in $V_1^{T=0}$ have negligible impact, and (b) the potential has a $\mathbb{Z}_2$ symmetry or that the breaking of this symmetry is sufficiently weak that the ${\bar b}_1\varphi$ term in Eq.~(\ref{eq:vsinglet1}) is negligible (for analytic ease we shall set it to zero in what follows). Under these assumptions, the equation $\partial V_{\rm eff}(\varphi,\alpha=\pi/2)/\partial\varphi=0$ has the following solutions:
\begin{eqnarray}
&&\varphi=0\\
&&\varphi_\pm=\frac{-{\bar b}_3\pm\sqrt{{\bar b}_3^2-4{\bar b}_2{\bar b}_4}}{2{\bar b}_4}\label{eq:phipm}
\end{eqnarray}
where the last solution is physical as long as $\Delta\equiv {\bar b}_3^2-4{\bar b}_2{\bar b}_4>0$. In what follows, we discuss the algebraic conditions that dictate which is the global minimum of the potential before EW symmetry breaking, i.e. whether or not $\varphi\not=0$ at $T>T_c$.

If $\Delta<0$ the origin is the high $T$ minimum, since $\partial^2 V_{\rm eff}(\varphi,\alpha=\pi/2)/\partial\varphi^2={\bar b}_2$, ${\bar b}_2>{\bar b}_3^2/(4{\bar b}_4)$ and ${\bar b}_4>0$. If $\Delta>0$ the origin is a maximum if ${\bar b}_2<0$ (case 1.), and a minimum if ${\bar b}_2>0$ (case 2.). 
\begin{itemize}
\item[1.] If ${\bar b}_2<0$, $\varphi_\pm$ are the two minima of the potential, and 
\begin{equation}\label{eq:veffbs}
V_{\rm eff}(\varphi_\pm,\alpha=\pi/2)=-\left(\frac{\bar b_2}{\bar b_4}+\frac{\bar b_3}{\bar b_4}\varphi_\pm\right)\left(\frac{\bar b_2}{4}-\frac{\bar b_3\varphi_\pm}{12}\right).
\end{equation}
To establish which is the true minimum, we compute
\begin{equation}
V_{\rm eff}(\varphi_+)-V_{\rm eff}(\varphi_-)=\frac{\sqrt{\Delta}}{6\bar b_4^2}\left(\frac{\bar b_3^2}{\bar b_4}-{\bar b}_2\right){\bar b}_3
\end{equation}
from which one reads (recalling that since $\Delta>0$ and $\bar b_4>0$ the term in parenthesis is positive) that if ${\bar b}_3>0$ the minimum is $\varphi_-$, while if ${\bar b}_3<0$ it is $\varphi_+$. 
\item[2.] If ${\bar b}_2>0$, the minimum is either the origin or $\varphi_+$ ($\varphi_-$) for $\bar b_3<0$ ( $\bar b_3>0$). The origin is the absolute minimum if $V_{\rm eff}(\varphi_\pm)>0$ {\em i.e.} (substituting (\ref{eq:phipm}) into (\ref{eq:veffbs})) if
\begin{equation}
6\bar b_2\bar b_4-\bar b_3^2>\mp \bar b_3\sqrt{\Delta}\qquad {\rm for}\ \varphi_\pm
\end{equation}
From the equation above, simply dividing right and left hand sides by $\bar b_3^2$, it follows that the origin is the absolute minimum if 
\begin{equation}\label{eq:R}
R\equiv\frac{\bar b_2\bar b_4}{\bar b_3^2} > \frac{2}{9}\ \ \ .
\end{equation}
In contrast, for $R<2/9$ and $\bar b_3>0$ the high-$T$ minimum is at $\varphi_-$, while if $R<2/9$ and $\bar b_3<0$ it is $\varphi_+$. 
\end{itemize}

In summary, in the limit where $\bar b_0=\bar b_1=0$, the global minimum of the potential before EW symmetry breaking lies at the origin if $R>2/9$ and at $\varphi\ne0$ if $R<2/9$.

When discussing the strength of the EWPT, it will be useful to know the magnitude of the high-$T$ minimum before EWSB. It is straightforward to show that for $0\leq R \leq 2/9$
\begin{eqnarray}
\label{eq:vsinglet2}
V_{\rm min}(\varphi_{\rm min},\alpha=\pi/2,T)\equiv V_0&=&-\frac{\bar b_3^4}{12\ \bar b_4^3}f(R)\\
\nonumber  f(R)&=&\frac{1}{2}\left[(1-4R)(1-2R)-2R^2+(1-4R)^{3/2}\right]\ \ \ .
\end{eqnarray}
The function $f(R)$ decreases monotonically from one to zero as $R$ is increased from zero to $2/9$. For $R>2/9, V_0=0$. Thus, if  $\bar b_3$ is of order  the electroweak scale or smaller and ${\bar b}_4\sim {\cal O}(1)$, then one would expect from Eq.~(\ref{eq:vsinglet2}) that $|V_0|/T^4 << 1$ for $T$ of order the electroweak scale. As we discuss below, our numerical study suggests that the critical temperature for the EWPT, $T_c$ is typically $\sim 100$ GeV for phenomenologically viable choices of the potential parameters. Thus, we will adopt an expansion in $|V_0|/T_c^4$ in analyzing the strength of the EWPT. Our result in Eq.~(\ref{eq:crit2}) relies on the validity of this expansion, that we have confirmed numerically.

The foregoing analysis can be generalized to the case where the linear term in $V_{\rm eff}(\varphi,\pi/2,T)$ is not negligible by shifting the classical field $\varphi$ by a $T$-dependent constant $\beta(T)$ as in Eq.~(\ref{eq:vshift}) to eliminate the ${\bar b}_1$ term and defining a new set of coefficients as in Eqs.~(\ref{eq:bshift}). The conditions for a nontrivial high-$T$ singlet vev that has $V_0<0$ will be the same as above, but expressed in terms of the shifted coefficients. The connection with the parameters in the tree-level potential will be implicit in these relations, which can be studied numerically. We have carried out such a numerical investigation of this issue, and we find that for models which do not exhibit a $\mathbb{Z}_2$ symmetry and that are compatible with LEP bounds on $(m_{k},\xi_{k})$ ($k=1,2$), one has $|V_0|/T_c^4 \ll 1$.

We checked the validity of our criterion outlined above against the full numerically computed finite temperature potential. We find that for almost all models with $R>2/9$ the origin is the absolute minimum at $T\gtrsim T_c$; however, we do find a few models with $0\lesssim R<2/9$ where the origin is the absolute minimum, indicating that $R<2/9$ is a necessary but not sufficient condition for the singlet field to develop a non-vanishing vev at $T\gtrsim T_c$. We also numerically compared the quantity $R$ in Eq.~(\ref{eq:R}) with the quantity $R_0=b_2b_4/b_3^2$ defined with the tree level couplings appearing in the potential, assuming a vanishing linear term $b_1$ in the potential, and we find that the deviation of $R_0$ from $R$ at $T=T_c$ is typically smaller than a few percent. Henceforth, for practical purposes, $R_0$ can be used to discriminate potentials where the singlet acquires a vev before the EWPT from those where it does not.

\section{The electroweak phase transition}
\label{sec:ewpt}

In order to explore the implications of the augmented scalar sector for the EWPT, it is convenient to express the effective potential for general values of $\alpha$ and $\varphi$ as 
\begin{equation}
\label{eq:vefffull}
V_{\rm eff}(\varphi,\alpha,T)=\bar\lambda\varphi^4+(e-\varepsilon T)\varphi^3+\left[2\bar D(T^2-T_0^2)+(b_2\sin^2\alpha)/2\right]\varphi^2+ BT^2\, \varphi +\cdots
\end{equation}
where
\begin{eqnarray}
\nonumber
B & = & \left(\frac{a_1+b_3}{12}\right)\, \sin\alpha\\
\nonumber
\bar D&=& D_{\rm SM}\cos^2\alpha+\frac{1}{48}\left[a_2(1+\sin^2\alpha)+3b_4\sin^2\alpha\right]\\
\nonumber
\varepsilon&=& 2\sqrt{2}E_{\rm SM}\cos^3\alpha\\
\label{eq:cubic}
e&=& \left(\frac{a_1}{2}\cos^2\alpha+\frac{b_3}{3}\sin^2\alpha\right)\sin\alpha\\
\nonumber
\bar\lambda&=& \bar\lambda_{\rm SM}(T)\cos^4\alpha+\frac{a_2}{2}\cos^2\alpha\sin^2\alpha+\frac{b_4}{4}\sin^4\alpha+\cdots\ \ \ ;
\end{eqnarray}
where $D_{\rm SM}$, $E_{\rm SM}$ and $\bar\lambda_{\rm SM}$ correspond to the usual SM values \cite{Quiros:1999jp}; and where the \lq\lq $+\cdots$" denote terms involving non-analytic dependence on $\varphi$ associated with the one-loop finite $T$-effects and the screened thermal masses. In principle, one could expand the latter in powers of $m^2/T^2$ as we did in Section \ref{sec:highT}, thereby obtaining additional contributions to ${\bar D}$, $B$, and ${\bar\lambda}$. Doing so here, however, is not particularly enlightening so we will simply work with the expression given in Eq.~(\ref{eq:vefffull}). 

In analyzing the parameters in the tree-level potential of Eq.~(\ref{eq:vtree}) that lead to a strong first order EWPT, we perform numerical scans while implementing the various physical considerations discussed above. In doing so, we utilize the full finite temperature effective potential of Eq.~(\ref{eq:vTeff}) without resorting to an expansion in $m^2/T^2$. In order to interpret the results, however, it is useful to carry out an analytic study, relying on approximations justified by our full numerical studies. To that end we first note that our numerical study indicates that the effects of the $BT^2\varphi$ term in Eq.~(\ref{eq:vefffull}) are relatively unimportant compared to the tree-level cubic term proportional to $e$. Although both $B$ and $e$ depend linearly on the parameters $a_1$ and $b_3$, the factor of $\sin\alpha/12$ in $B$ suppresses their effect, especially when the value of $\sin\alpha$ is small near the critical temperature, as we find is typically the case. Thus, dropping the linear term in Eq.~(\ref{eq:vefffull}) is a well-justified and helpful approximation.

We now derive conditions on the parameters such that the condition for a strong first order EWPT (following from Eq.~(\ref{eq:crit1}) and making use of the projection of $v(T_c)$ over the SU(2)$_L$ direction \cite{extrasinglet,othersinglet,Pietroni:1992in,Ahriche:2007jp})
\be
\frac{v(T_c)}{T_c}=\sqrt{2}\cos\alpha_c\, \left(\frac{\varphi_c}{T_c}\right)\gtrsim 1
\ee
obtains. The values of $\varphi_c$, $\alpha_c$, and $T_c$ are determined from the two minimization conditions
\begin{eqnarray}
\label{eq:minphi}
\frac{\partial V_{\rm eff}(\varphi_{\rm min},\alpha_{\rm min},T)}{\partial\varphi} &=& 0\\
\label{eq:minalpha}
\frac{\partial V_{\rm eff}(\varphi_{\rm min},\alpha_{\rm min},T)}{\partial\alpha} & = & 0
\end{eqnarray}
and the condition that defines $T_c$, namely, that the potential at the minima of broken and unbroken electroweak symmetry as the same value:
\be
\label{eq:tc}
 V_{\rm eff}(\varphi_{\rm min},\alpha_{\rm min}\not=\pi/2,T_c)=V_{\rm eff}(\varphi_{\rm min}, \pi/2
 ,T_c)\equiv V_0\ \ \ ,
\ee
where $V_0$ is the high-$T$ minimum defined in Section \ref{sec:highT}. 
\FIGURE[!t]{\label{fig:potential1}
\mbox{\hspace*{-.8cm}\epsfig{file=FIGURES/potential1.eps,width=7.5cm}\qquad
\epsfig{file=FIGURES/phimin1.eps,width=8.02cm}}\\[0.7cm]
\mbox{\hspace*{-.8cm}\epsfig{file=FIGURES/iso1_2.eps,width=7.5cm}\qquad\qquad
\epsfig{file=FIGURES/iso1_1.eps,width=7.5cm}}\\
\caption{Upper panels: The potential $V_{\rm eff}(\varphi_{\rm min}(\alpha),\alpha,T)$ (left) and the location of the minima in the $(v/\sqrt{2},x)$ plane (right) for a model featuring the origin as its high-T vacuum, for various values of the temperature $T$. Lower panels: A contour map showing iso-level curves of constant values of the potential at $T=65$ GeV (left) and at $T=0$ (right). The red arrows point towards smaller values of the potential. The black dots indicates the location of the absolute minimum of the potential.}}
\FIGURE[!t]{\label{fig:potential2}
\mbox{\hspace*{-.8cm}\epsfig{file=FIGURES/potential2.eps,width=7.5cm}\qquad
\epsfig{file=FIGURES/phimin2.eps,width=8.02cm}}\\[0.7cm]
\mbox{\hspace*{-.8cm}\epsfig{file=FIGURES/iso2_2.eps,width=7.5cm}\qquad\qquad
\epsfig{file=FIGURES/iso2_1.eps,width=7.5cm}}\\
\caption{Upper panels: The potential $V_{\rm eff}(\varphi_{\rm min}(\alpha),\alpha,T)$ (left) and the location of the minima in the $(v/\sqrt{2},x)$ plane (right) for a model featuring a non-zero singlet vev at $T\gtrsim T_c$, for various values of the temperature $T$. Lower panels: A contour map showing iso-level curves of constant values of the potential at $T=340$ GeV (left) and at $T=0$ (right). The red arrows point towards smaller values of the potential. The black dots indicates the location of the absolute minimum of the potential.}}

This procedure is illustrated in Fig.~\ref{fig:potential1} and \ref{fig:potential2}. The model we consider in Fig.~\ref{fig:potential1} does not develop a singlet vev prior to the EWPT, and is characterized by the parameter set $(a_1,a_2,b_3,b_4,x_0,\lambda)=(-933,0.69,356,0.53,55,0.4)$, with the parameters having mass dimension 1 in units of GeV. The model we study in Fig.~\ref{fig:potential2} features instead a non-zero singlet vev at $T\gtrsim T_c$, and is defined by $(a_1,a_2,b_3,b_4,x_0,\lambda)=(-88,0.765,-100,0.46,300,0.3)$. In the upper left panels we show the value of the potential as a function of the angle $\alpha$ at the location of the minimum in the radial $\varphi$ direction, defined by Eq.~(\ref{eq:minphi}). In the upper right panels, we indicate the location of the corresponding valleys of minima on the $(v/\sqrt{2},x)$ plane. The two lower panels show iso-level contour maps illustrating the behavior of the potential in the $(v/\sqrt{2},x)$ plane at high temperature and at $T=0$. The red arrows point towards smaller values of the potential. The black dots indicates the location of the absolute minima.

Let us first consider Fig.~\ref{fig:potential1}. From the upper left panel we see that when $T=65$ GeV, for $\alpha/\pi\lesssim0.062$ the potential does not develop a minimum in the radial direction, hence the discontinuity in the resulting curve. The absolute minimum is at the origin, as indicated by the black dot in the right panel. Following again the curve at $T=65$ GeV, we see that for $0.062\lesssim\alpha/\pi\lesssim0.13$ there exist minima in the radial direction, whose location is shown by the double-dotted-dashed line in the right panel. However, the values of the potential along that line are always larger than the value at the origin, as indicated by the same line in the left panel. Therefore, the absolute minimum is at the origin (the black dot in the right panel). Lowering the temperature, the potential on the line of radial minima decreases, until, at $T=T_c$ (dot-dashed line) its value at $\alpha=\alpha_c\simeq0.093\pi$ equals the value at the origin. In the $(v/\sqrt{2},x)$ plane, upper right, the corresponding point at $\varphi_{\rm min}(\alpha_c)$ is shown by the black dot on the dot-dashed line. Finally, the black line, corresponding to $T=0$, indicates that the minima in the radial direction exist now in the range $0<\alpha/\pi\lesssim0.13$, and that the vacuum of the theory corresponds to $\alpha_0\simeq0.95$ and to the black dot on the solid black curve in the right panel on the $(v/\sqrt{2},x)$ plane. As explained above, the two lower panels help figuring out the shape of the potential at $T=65$ GeV (left) and at $T=0$ (right); the black dots indicate the location of the absolute minimum of the potential.

Similarly, in Fig.~\ref{fig:potential2}, the minimum of the potential at $T=137$ GeV is at $x\simeq385$ GeV and $v=0$, i.e. at $\alpha/\pi=0.5$, as indicated in the left panel. At the critical temperature (dashed line in the left panel) the value of the potential at $\alpha_c/\pi\simeq0.34$ equals that at $\alpha/\pi\simeq0.5$. The location of that minimum in the $(v/\sqrt{2},x)$ plane is shown by the black dot on the solid black line in the right panel. For temperatures below $T_c$ the location of the minima in the radial direction doesn't change visibly on the $(v/\sqrt{2},x)$ plane, but the value of the potential at the true vacuum keeps decreasing, as indicated by the solid black line in the left panel. 

We now solve Eqs.~(\ref{eq:minphi}) and (\ref{eq:tc}) analytically and study the resulting potential $V_{\rm eff}[\varphi_{\rm min}(\alpha), \alpha, T_c(\alpha)]$ as a function of $\alpha$ as above to find the minima of the potential. It is useful to consider first the case for $V_0=0$. The solution to Eqs.~(\ref{eq:minphi}) and (\ref{eq:tc}) then yields
\be
\label{eq:phitc0}
\left(\frac{\varphi_c}{T_c}\right)_0 = \frac{\varepsilon-e/T_c}{2{\bar\lambda}}\ \  ,
\ee
where the \lq\lq $0$" subscript denotes case for $V_0=0$. As indicated by Eqs.~(\ref{eq:cubic}), the values of $\varepsilon$, $e$, and $\bar\lambda$  are explicit functions of $\alpha$ while $T_c$ depends on $\alpha$ implicitly. The value of $\alpha=\alpha_c\not=\pi/2$ that minimizes the potential is obtained from a numerical examination of $V_{\rm eff}[\varphi_{\rm min}(\alpha), \alpha, T_c(\alpha)]$ and determines the value of the ratio in Eq.~(\ref{eq:phitc0}), as described in the discussion of Fig.~\ref{fig:potential1}. In order to prevent washout of the baryon asymmetry by electroweak sphalerons, the projection of $\varphi_c/T_c$ along the SU(2)$_L$ direction must be $\gtrsim 1$ \cite{Ahriche:2007jp},
\be
\label{eq:phicond}
\sqrt{2}\cos\alpha_c\, \left(\frac{\varphi_c}{T_c}\right)\gtrsim 1
\ee
leading to the result anticipated in Eq.~(\ref{eq:crit2}). 

We analyze the case when $V_0\not=0$ by recalling from the analysis of Section \ref{sec:highT} that one has $|V_0|/T_c^4 << 1$ for phenomenologically realistic parameters in the singlet potential. Thus, we solve Eqs.~(\ref{eq:minphi}) and (\ref{eq:tc}) by expanding in powers of $|V_0|/T_c^4$ about the $V_0=0$ solution and obtain, to leading non-trivial order in $|V_0|/T_c^4$,
\be
\label{eq:phitc1}
\left(\frac{\varphi_c}{T_c}\right)=\left(\frac{\varphi_c}{T_c}\right)_0\, \left[1+\gamma\frac{|V_0|}{T_c^4}\right] +\cdots \ \ \ ,
\ee
where the $+\cdots$ indicate higher order terms in the  $|V_0|/T_c^4$ expansion and where 
\be
\label{eq:gammadef}
\gamma = \frac{[3\varepsilon-2{\bar\lambda}(\varphi_c/T_c)_0][\varepsilon-e/T_c]-16{\bar\lambda}{\bar D}}{2\varepsilon(\varepsilon-e/T_c)-16{\bar\lambda}{\bar D}}\, \left(\frac{\varphi_c}{T_c}\right)_0^{-4}\ \ \ .
\ee
Here, the value $T_c$ appearing in Eqs.~(\ref{eq:phitc1},\ref{eq:gammadef}) is the value  obtained for $V_0=0$. 
From Eqs.~(\ref{eq:phicond}) and (\ref{eq:phitc1}) we obtain the result in Eq.~(\ref{eq:crit2}).

It is relatively straightforward to demonstrate that the coefficient $\gamma$ is positive for realistic parameter choices\footnote{For instance, in the SM limit $16\bar\lambda\bar D\gg\varepsilon^2$.}, so that the presence of a non-trivial high-$T$ minimum always strengthens the EWPT. We have explicitly verified and confirmed this analytic result with our full numerical study. Thus, for practical purposes, it suffices to study Eq.~(\ref{eq:crit2}) for $V_0=0$ in order to determine the potential parameters for which one obtains a sufficiently strong first order EWPT. Before doing so numerically, it is instructive to express this condition in terms of the parameters in the potential:
\be
\label{eq:phitc2}
\frac{4E_{SM}\cos^4\alpha_c-\sqrt{2}\cos\alpha_c\sin\alpha_c\left(a_1\cos^2\alpha_c/2+b_3\sin^2\alpha_c/3\right)T_c}{2{\bar\lambda}_0(T_c)\cos^4\alpha_c+a_2\cos^2\alpha_c\sin^2\alpha_c+b_4\sin^4\alpha_c/4} 
\gtrsim 1 \ \ \ ,
\ee
where we have neglected the non-analytic dependence on the background field as before. 

We now make several observations about  Eq.~(\ref{eq:phitc2}). 
\begin{itemize}
\item[(i)] In the limit of  pure SU(2)$_L$ scalar sector, one has $\alpha_c=0$ and Eq.~(\ref{eq:phitc2}) reproduces the SM result of Eq.~(\ref{eq:critsm}) with the substitution $\bar\lambda_0(T_c)\approx {\bar\lambda}_0(0)= m_h^2/2v_0^2$. 
\item[(ii)] The introduction of a singlet extension of the scalar sector can lead to an increase in the strength of the EWPT in one of four ways\footnote{As noted in the introduction, choosing the parameters to give $V_0<0$ leads to a fifth avenue for strengthening the EWPT; the conditions for achieving this situation have been discussed in Section \ref{sec:highT}.}: (1) the presence of the terms in the numerator of Eq.~(\ref{eq:phitc2}) associated with the breaking of $\mathbb{Z}_2$ symmetry, so long as $\cos\alpha_c\sin\alpha_c (a_1\cos^2\alpha_c/2+b_3)<0$; (2) a reduction in the magnitude of the denominator due to a negative value for $a_2$ (but still satisfying the condition that the potential is bounded from below); (3) a reduction of the denominator due to  singlet contributions to the one-loop effective potential that renormalize $\lambda_0$ \cite{Anderson:1991zb}; and (4) singlet contributions to the parameter $\varepsilon$ that we have not shown explicitly. 
\item[(iii)] Singlet contributions to $\varepsilon$ occur under specialized choices of the singlet parameters that lead to a cancellation of the mass functions $\Pi$ as in the MSSM case with the light scalar top, leading to an additional $\varphi^3 T$ term in the potential beyond that of the SM. This special case has been studied in Ref.~\cite{Espinosa:2007qk} for models having $\mathbb{Z}_2$ symmetry, and it was found that introduction of a large number of singlet scalars [$\sim 10$] is needed to bring about a sufficiently strong first order EWPT for realistic masses for the SM-like Higgs boson. 
\item[(iv)] The effect of singlet renormalization of $\bar\lambda_0$ as studied in Ref.~\cite{Anderson:1991zb} is suppressed by a factor of $1/64\pi^2$, so a large value of the parameter $a_2$ that enters this renormalization is needed to bring about an appreciable effect. Since the value of $a_2$ contributes to the degree of mixing between the singlet-like and the SM-like mass eigenstates and also governs the magnitude of $\Gamma(h_i\to h_j h_j)$, a value of $|a_2|$ that is large enough to produce substantial renormalization of $\bar\lambda_0$ will have noticeable phenomenological consequences (see  Sec.~\ref{sec:pheno}). The most dramatic effect, when $m_1>2m_2$, is a large suppression of the branching ratio for the decay of the lightest scalar to ordinary SM decay modes, as the largest decay width will correspond to a pair of singlet-like scalars.
\item[(v)] In contrast, mechanisms (1) and (2) require no special tuning of parameters to yield a sufficiently strong first order EWPT when only one light scalar singlet is present. Not surprisingly, most of the models we find in our numerical scan correspond to these two cases.
\end{itemize}

\FIGURE[]{\label{fig:critalpha}
\epsfig{file=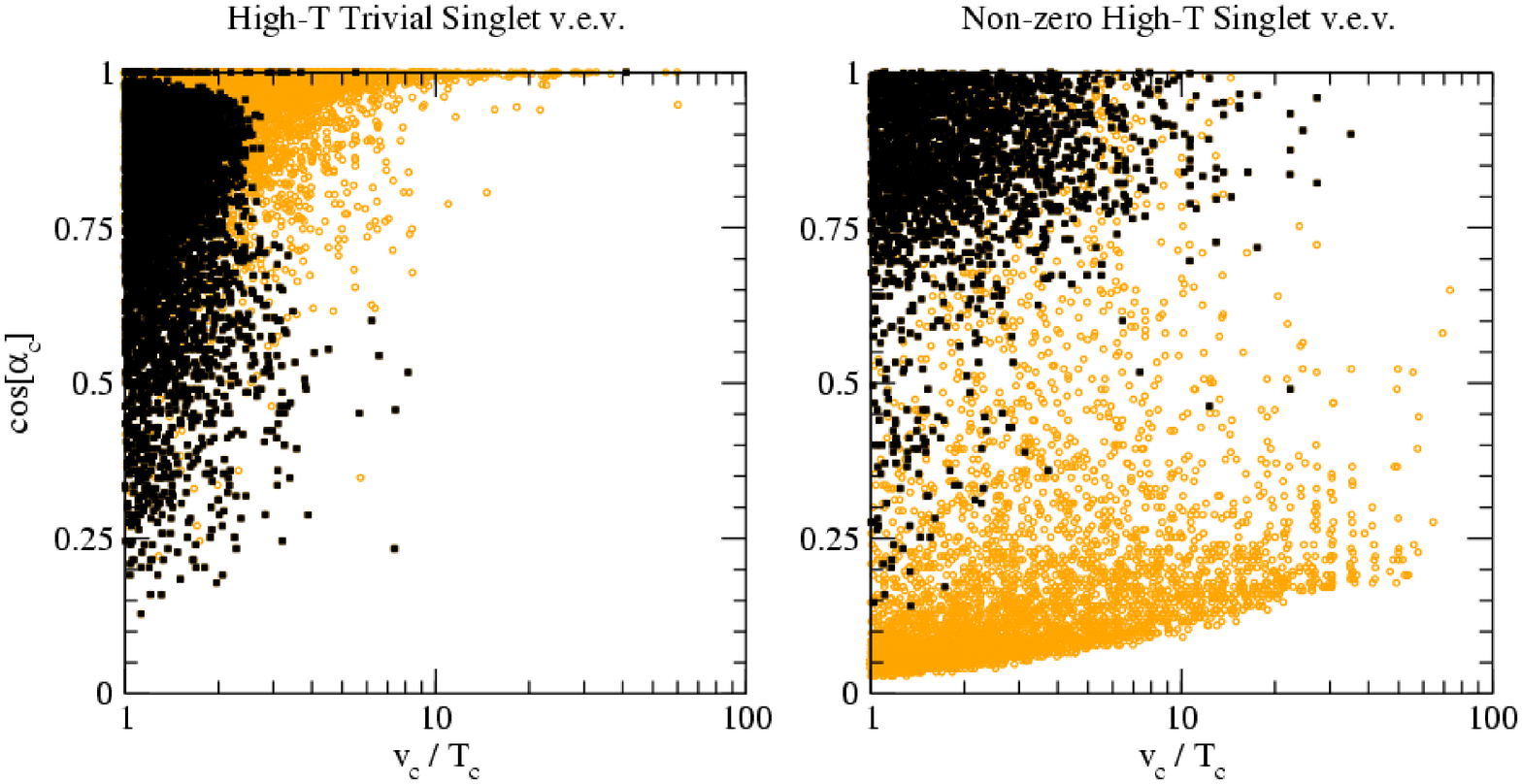,width=15cm}
\caption{The correlation between the order parameter $v_c/T_c$ and the critical temperature angle $\alpha_c$ for models featuring a strongly first order EWPT for the parameter space scan outlined in Tab.~\ref{tab:param}. Black square points correspond to models also consistent with Higgs searches at LEP 2. The left panel shows models with a vanishing singlet vev at $T\gtrsim T_c$, while that to the right models with a non-zero singlet vev at $T\gtrsim T_c$.}}
Before proceeding to low-energy probes of the parameters relevant to the EWPT, we consider Eq.~(\ref{eq:phitc2}) in the limit of small $\sin\alpha_c$. As indicated in Fig. \ref{fig:critalpha}, a strong first order EWPT occurs quite readily for small $\sin\alpha_c$, especially in the case where the singlet acquires a vev before the EWPT (right panel). Even in the left panel we find that a large fraction of points have $\cos\alpha_c\sim1$, although this is hardly visible and quantifiable from the figure, where the points are all aligned in the upper part of the plot. 

Retaining only terms to linear order in $\sin\alpha_c$ we obtain from Eq.~(\ref{eq:phitc2})
\be
\label{eq:smallalpha1}
\frac{2 E_{SM}}{\bar\lambda_0} -\frac{1}{2\sqrt{2}}\, \frac{a_1}{T_c}\, \frac{\tan\alpha_c}{\bar\lambda_0}\gtrsim 1\ \ \ .
\ee
Thus, one sees that in this regime, a relatively large negative value of $a_1/T_c$ is needed for substantial enhancements of the EWPT compared to the SM situation. From Eqs.~(\ref{eq:massparams}-\ref{eq:mixing}), $a_1$ also governs the degree of mixing between the singlet and SU(2)$_L$ neutral scalar in the case of small $a_2$. Thus, for sufficiently light singlets, Higgsstrahlung studies at $e^+e^-$ colliders can probe the parameter space needed for a strong first order EWPT. To see this situation explicitly, we consider the regime in which $\mu_s^2<< \mu_h^2$ and express the LHS of Eq.~(\ref{eq:smallalpha1}) in terms of the mixing angle $\theta$ and neutral SU(2)$_L$ scalar mass:
\be
\label{eq:smallalpha2}
4E_{SM}\, \left(\frac{v_0^2}{\mu_h^2}\right) -\left(\frac{2\tan\theta\tan\alpha_c}{1-\tan^2\theta}\right)\, \left(\frac{v_0}{T_c}\right)\, \gtrsim 1\ \ \ .
\ee
For light $\mu_s^2$, large values of $\tan\theta$ are excluded by LEP searches for a light scalar with reduced couplings to the $Z^0$ boson. Consequently, we observe in Fig. \ref{fig:critalpha} that imposing the LEP constraints significantly reduces the number of parameter space points that lead to a strong first order EWPT. In Section \ref{sec:pheno}, we explore the implications of present and prospective Higgsstrahlung studies without resorting to the small $\sin\alpha_c$ and $\mu_h^2 >> \mu_s^2$ limit. Nevertheless, the limiting case given in Eqs.~(\ref{eq:smallalpha1},\ref{eq:smallalpha2}) provides useful guidance for interpreting the more general numerical studies. 

\FIGURE[]{\label{fig:xlambda}
\epsfig{file=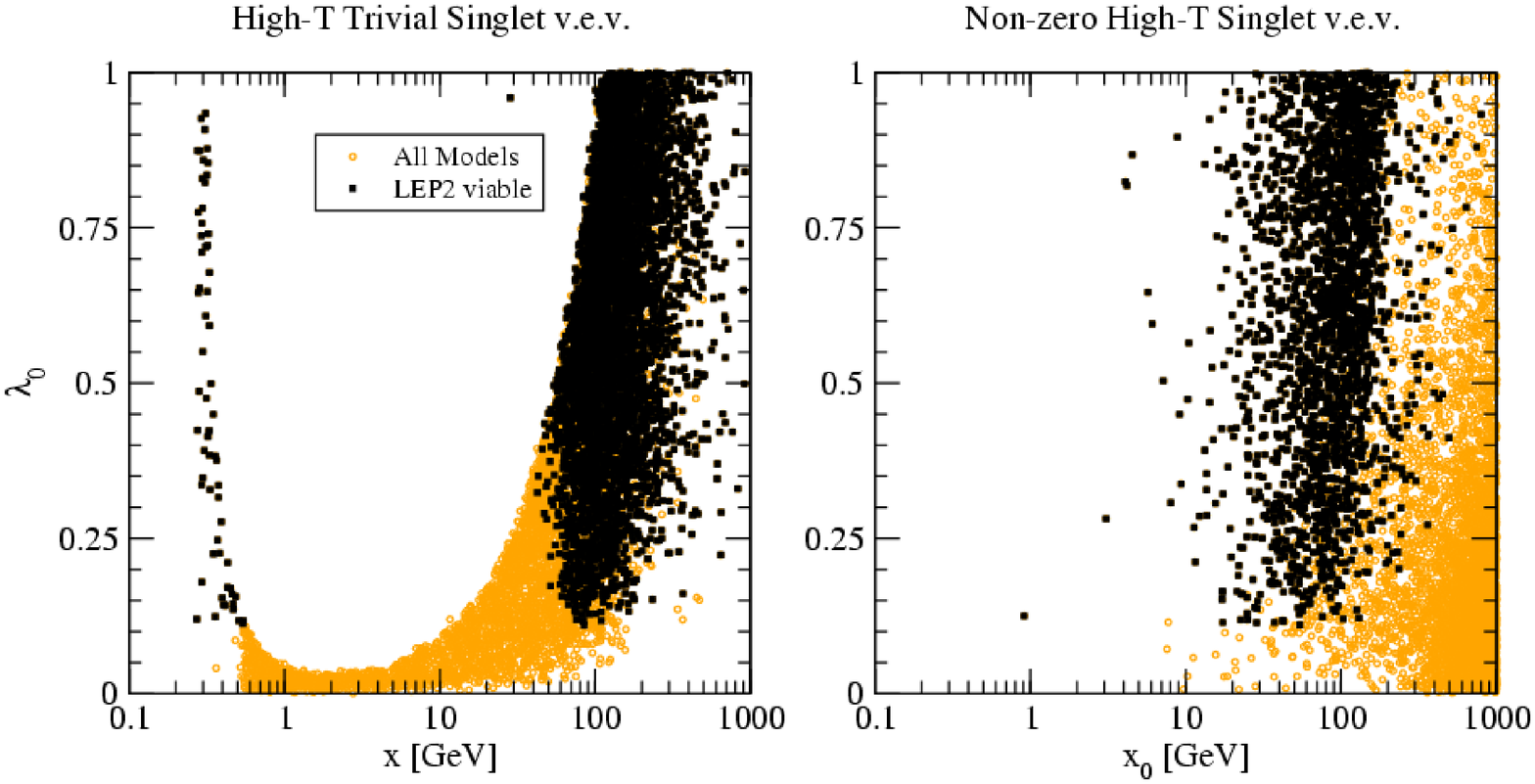,width=15cm}
\caption{As in Fig.~\ref{fig:critalpha}, for $x_0$ and $\lambda_0(T=0)$.}}

It is also instructive to analyze the dependence of the LHS of the inequality (\ref{eq:phicond}) on the parameters in the tree-level potential. In Fig. \ref{fig:xlambda}, we show the values of the zero temperature singlet vev $x_0$ and SU(2)$_L$ quartic coupling $\lambda_0(0)$ that lead to a strong first order EWPT. The points in Fig.  \ref{fig:xlambda} were generated by scanning over the potential parameters as discussed above, and requiring a strongly first order EWPT; points in black are consistent with the LEP Higgs search constraints on the $(m_{k},\xi_{k}^2)$ plane ($k=1,2$), while the black and light colored points together indicate the results of our scan over all models. The results are shown for the case when both the high-$T$ singlet vev is zero (left panel) or non-vanishing with $V_0<0$ (right panel). We  observe that models with small values of $\lambda_0(T=0)$ are excluded by LEP since they lead to exceedingly small values for the mass of the SM-like neutral scalar [see Eq.~(\ref{eq:massparams})]. Moreover, models having a substantial zero-temperature singlet vev and a strong first order EWPT occur quite abundantly, even after the LEP constraints are implemented. We also notice that, unlike the case where the singlet acquires a vev before the EWPT, when the EWPT proceeds from zero vev for the singlet (left panel) various models feature very small values for $x_0$ and are LEP-compatible (the vertical string of points in the left part of the plot). For those models, $\mu_s^2$ is typically very large due to the $a_1v_0^2/(4x_0)$ term in Eq.\ref{eq:massparams}, and the mixing between the two scalars is suppressed

\FIGURE[]{\label{fig:a1a2}
\epsfig{file=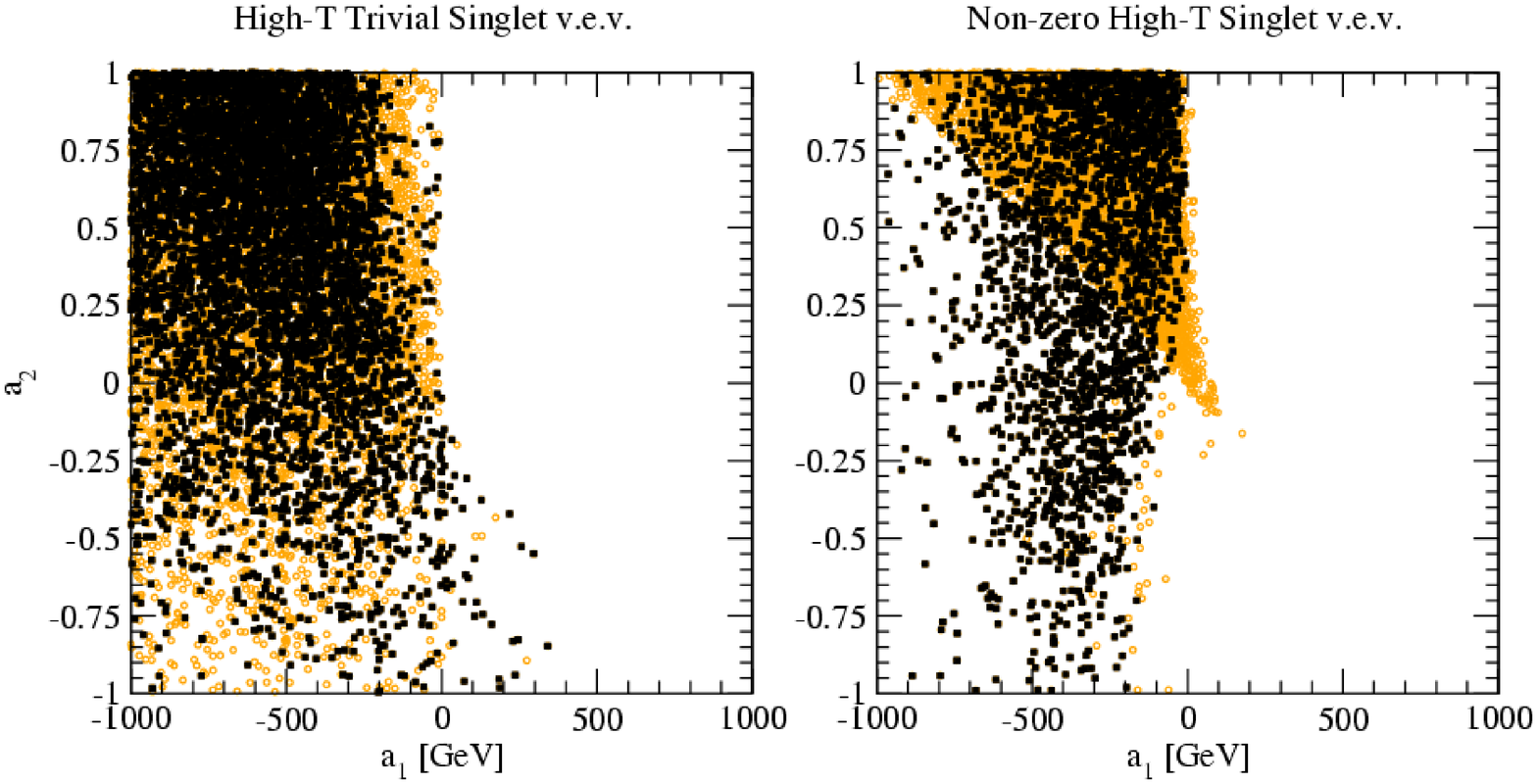,width=15cm}
\caption{As in Fig.~\ref{fig:critalpha}, for the mixing parameters $a_1$ and $a_2$.}}
\FIGURE[!t]{\label{fig:sms2}
\mbox{\epsfig{file=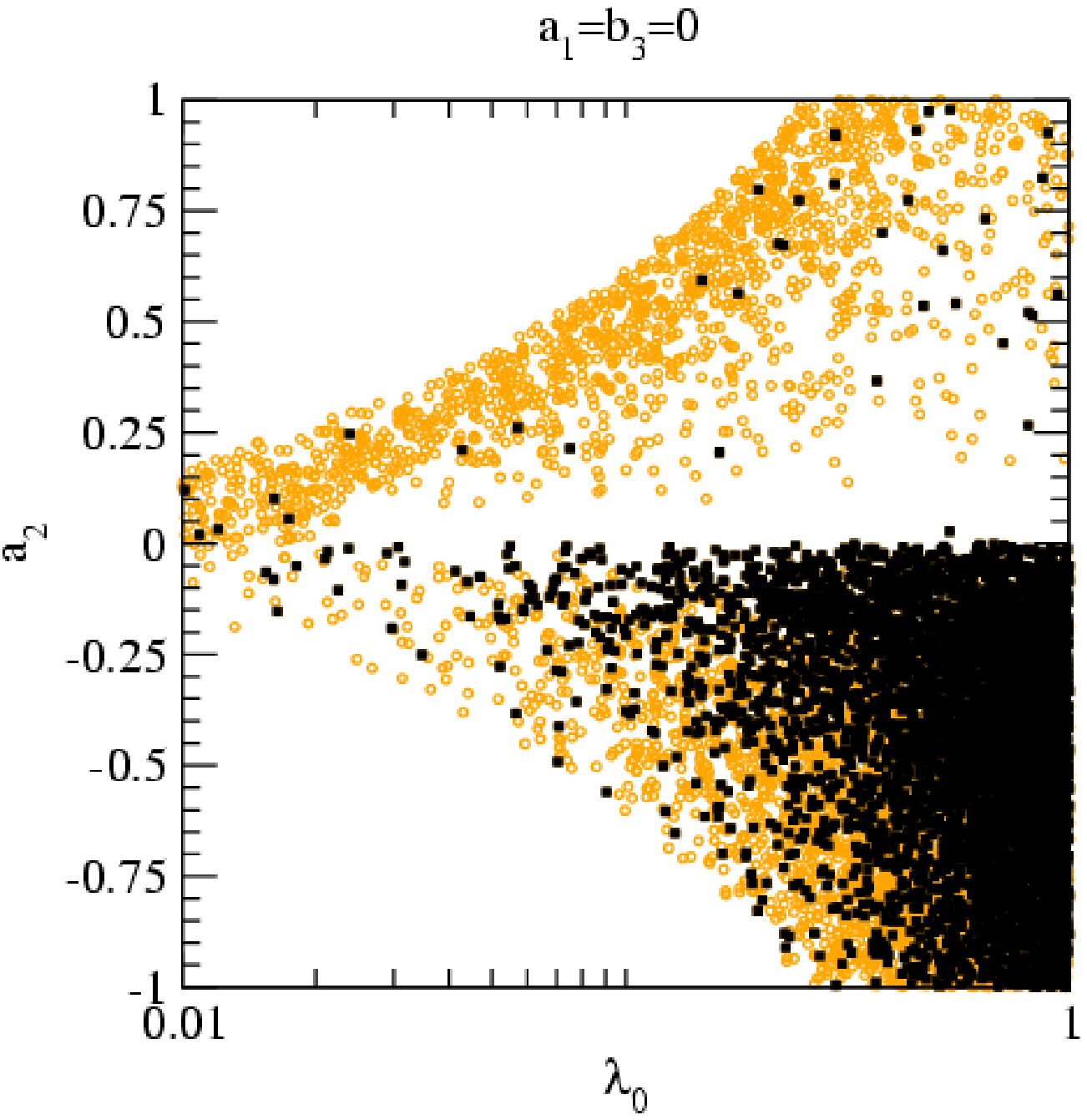,width=7.7cm}\qquad
\epsfig{file=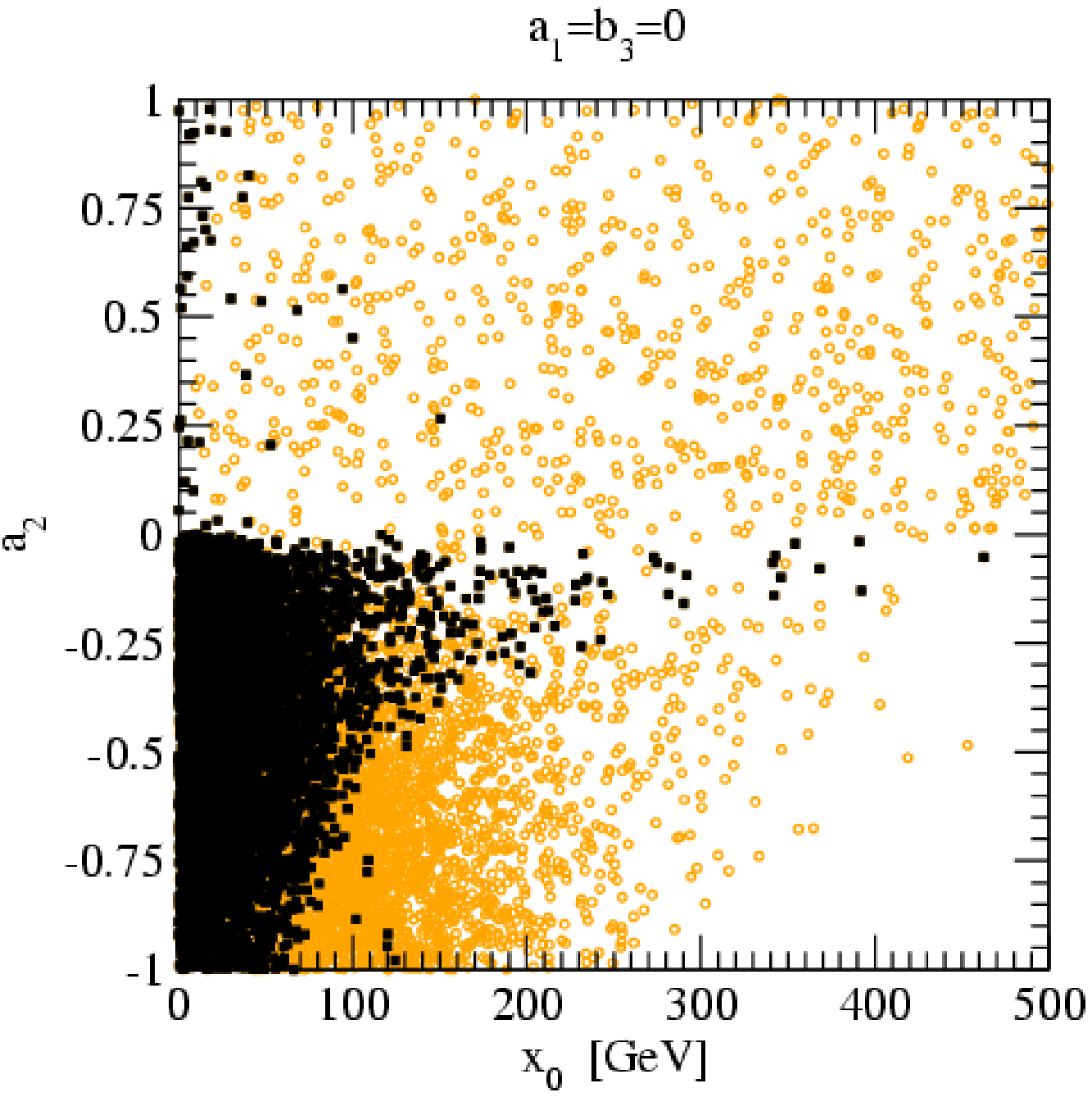,width=7.7cm}}
\caption{The $(\lambda,a_2)$ and $(x,a_2)$ planes for models with $a_1=b_3=0$.}}
In Figure \ref{fig:a1a2} we show the distribution of parameters $a_1$ and $a_2$ that couple the SU(2)$_L$ and singlet Higgs and that lead to a strong first order EWPT. We note that in nearly all cases, the $\mathbb{Z}_2$ symmetry-breaking parameter $a_1$ is negative, as one expects from the discussion of Eq.~(\ref{eq:phitc2}). Models with $a_2$ of either sign can lead to a strong first order EWPT whether or not the LEP constraints are implemented. Note that when $|a_1|$ is large, a positive value for $a_2$ is more likely to be consistent with the LEP constraints. This trend can be understood qualitatively from the dependence of $\mu_{hs}^2$ on these two parameters; to avoid large mixing between the singlet and SU(2)$_L$ neutral scalars, the value of $a_2$ must be relatively large and positive to cancel a large negative contribution to $\mu^2_{hs}$ from a large negative $a_1$ when $x_0$ differs significantly from zero. When both $a_1$ and $a_2$ are negative, however, the constraints on large mixing can be relaxed by making $\mu_s^2$ large by suitable values of $b_{3,4}$, a situation corresponding to the points in the lower left quadrants of both panels in Fig. \ref{fig:a1a2}. We also find that in models with $\mathbb{Z}_2$ symmetry, a strong first order EWPT is generally consistent with the LEP constraints only when $a_2<0$, as indicated in Fig. \ref{fig:sms2}. Again, this behavior is consistent with Eq.~(\ref{eq:phitc2}) since a negative value of $a_2$ can reduce the denominator of LHS without engendering large singlet-doublet mixing.

\FIGURE[]{\label{fig:phivstc}
\epsfig{file=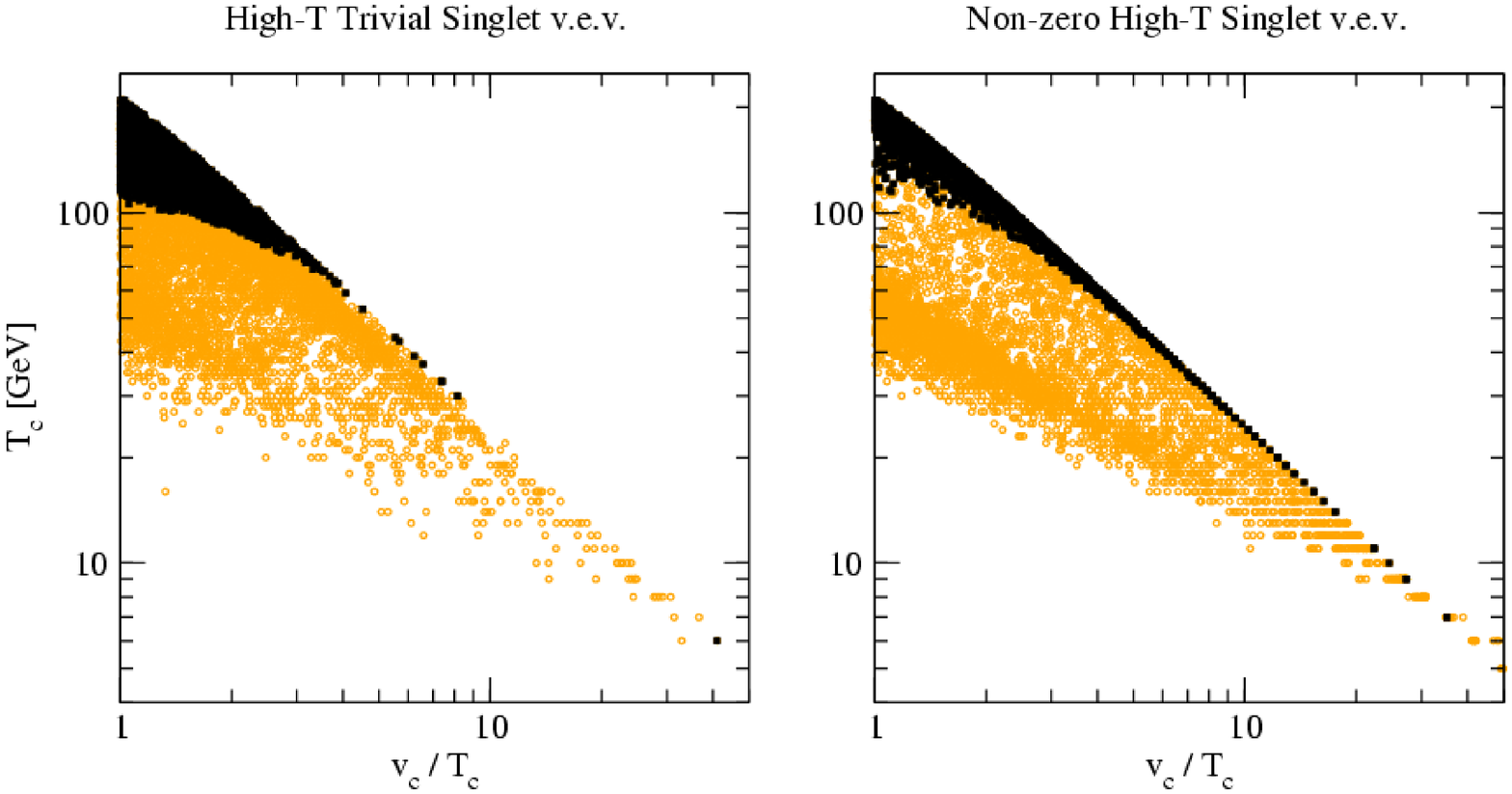,width=15cm}
\caption{As in Fig.~\ref{fig:critalpha}, for the order parameter versus the critical temperature.}}
Finally, we show in Fig. \ref{fig:phivstc} the dependence of $\sqrt{2}\cos\alpha_c\varphi_c/T_c=v_c/T_c$ versus $T_c$. Here, we see that imposition of the LEP constraints has a substantial impact. In the absence of such constraints, the value of $T_c$ leading to a strong first order EWPT can vary from as much as $\sim 200$ GeV to $10$ GeV, whereas imposing the LEP bounds favors critical temperatures of order $100$ GeV. Hence, for the $V_0\neq 0$ case, our expansion in $|V_0|/T_c^4$ is good for the LEP-allowed models.

\section{The zero-temperature phenomenology}
\label{sec:pheno}

As the discussion of Section \ref{sec:ewpt} indicates, searches for the Higgs boson and studies of its properties can provide important information about the viability of a strong first order EWPT in the presence of an extended scalar sector. Here, we discuss the present implications of EWPO as well as the possible tests of this scenario that might be carried out at the LHC and at a Linear Collider.

\vskip 0.1in
\noindent{\em Electroweak Precision Observables \& Singlet Scalars}
\vskip 0.1in

We first consider the impact of EWPO on singlet models that also lead to a strong first order EWPT. We address this issue by computing the scalar contributions to the gauge boson propagators (Fig.~\ref{fig:feyn}) in the Feynman-'t Hooft gauge and compare the corresponding effects with those from the SM Higgs on EWPO. Although the masses $m_{1,2}$ are generally of order the electroweak scale, it is possible to characterize their effects on the $Z^0$-pole observables and $W$-boson mass with the oblique parameters $S$, $T$, $U$, and $V$. The effects on lower-energy precision observables, such as the weak charge of the cesium nucleus obtained from atomic parity violation \cite{Wood:1997zq} or the parity-violating asymmetry in M\o ller scattering \cite{Anthony:2003ub} require that one take into account the $q^2$-dependence of the gauge boson self energies that is often approximated using the additional oblique parameters $W$, $X$, {\em etc.}. In the present analysis, we will neglect these higher derivative terms, saving a more complete analysis to a future publication \cite{radiativeinprep}. Note that the introduction of a real neutral scalar does not affect the propagator functions $\Pi_{\gamma\gamma}$ and $\Pi_{Z\gamma}$, so it suffices to consider only the effects on $\Pi_{WW}$ and $\Pi_{ZZ}$ in the present situation. 

\FIGURE[]{\label{fig:feyn}
  \begin{picture}(300,100)(0,0)
    \Photon(0,30)(30,30){5}{3}
    \Photon(60,30)(90,30){5}{3}
    \DashCArc(45,30)(15,-360,0){3}
    \Photon(110,30)(170,30){5}{6}
    \DashCArc(140,50)(15,-360,0){3}
    \Photon(190,30)(220,30){5}{3}
    \Photon(250,30)(280,30){5}{3}
    \DashCArc(235,30)(15,0,180){3}
    \PhotonArc(235,30)(15,180,0){3}{5}
    \Text(35,50)[lb]{{{$h,\ S$}}}
    \Text(10,40)[lb]{{{$V$}}}
  \end{picture}
\caption{Feynman diagrams for scalar one-loop corrections to the gauge boson propagators.}}
A simple indication of the difference between the effects of the SM Higgs and those of the mixed states $h_{1,2}$ on the precision observables can be seen from the expressions for scalar contributions to the $T$ parameter in the doublet-singlet scenario as compared to the SM case:
\begin{eqnarray}
\nonumber
T-T_{SM} & = & \left(\frac{3}{16\pi{\hat s}^2}\right)\Biggl\{ \frac{1}{c^2}\left(\frac{m_h^2}{m_h^2-M_Z^2}\right)\, \ln\, \frac{m_h^2}{M_Z^2} - \left(\frac{m_h^2}{m_h^2-M_W^2}\right)\, \ln\, \frac{m_h^2}{M_W^2}\\
\label{eq:deltaT}
& - & \cos^2\theta\, \left[\frac{1}{c^2}\left(\frac{m_1^2}{m_1^2-M_Z^2}\right)\, \ln\, \frac{m_1^2}{M_Z^2} - \left(\frac{m_1^2}{m_1^2-M_W^2}\right)\, \ln\, \frac{m_1^2}{M_W^2}\right]\\
\nonumber
& - & \sin^2\theta\, \left[\frac{1}{c^2}\left(\frac{m_2^2}{m_2^2-M_Z^2}\right)\, \ln\, \frac{m_2^2}{M_Z^2} - \left(\frac{m_2^2}{m_2^2-M_W^2}\right)\, \ln\, \frac{m_2^2}{M_W^2}\right]\Biggr\}\ \ \ ,
\end{eqnarray}
where ${\hat s}^2\equiv\sin^2{\hat\theta}_W(M_Z)$ gives the weak mixing angle in the $\overline{MS}$ scheme at the t'Hooft scale $\mu=M_Z$, $c^2=M_W^2/M_Z^2$, and $m_h$ is the reference value of the Higgs boson mass in the SM.  Note that the effect of the singlet enters both through the presence of the mixing angles in the prefactors as well as the values of the masses $m_{1,2}$ in the expressions that they multiply. The expressions for $S-S_{SM}$ and $U-U_{SM}$ are analogous, but too cumbersome to reproduce here since they involve evaluation of the loop integral functions 
\be
F_n(m_a^2, m_b^2, q^2) \equiv \int_0^1\, dx\, x^n\, \ln\left[(1-x)m_a^2+x m_b^2-x(1-x)q^2\right]\ \ \ .
\ee

In order to identify the EWPT-viable models that are favored by EWPO, we have performed a fit to the precision observables identified in the Review of Particle Physics (RPP) \cite{Yao:2006px} for a fixed value of $m_h=114.4$ GeV and obtained the values of $\Delta O\equiv O-O_{SM}$, $O=S$, $T$, $U$. To carry out this fit, we have employed the GAPP program \cite{gapp} that has been used in arriving at the best-fit values in the RPP, but using the value of the top quark mass $m_t=170.9\pm1.8$ given in Ref.~\cite{unknown:2007bx}. The results give the best fit values and standard errors\footnote{The GAPP routine includes SM Higgs radiative corrections explicitly. Consequently, the SM Higgs contributions to the oblique parameters for a fixed value of $m_h$ must be subtracted from the singlet-doublet scalar contributions when interpreting the results of the fit.}
\begin{eqnarray}
\nonumber
T-T_{SM} & = & -0.111\pm 0.109\\
\label{eq:obliquefit}
S-S_{SM} & = &-0.126\pm 0.096 \\
\nonumber
U-U_{SM} & = & 0.164\pm 0.115
\end{eqnarray}
We define a $\Delta\chi^2$ {\em via}
\be\label{eq:deltachi2}
\Delta\chi^2=\sum_{i,j} (\Delta O_i-\Delta O_i^0) \left(\sigma^{2}\right)^{-1}_{ij} \left(\Delta O_j-\Delta O_j^0\right) \ \ \ ,
\ee
where $\Delta O_i^0$ denotes the best fit value for the differences in a given oblique parameter from its SM reference value; $\sigma^2_{ij}=\sigma_i\rho_{ij}\sigma_j$, with $\sigma_i$ denoting the errors in Eq.~(\ref{eq:obliquefit}); and the correlation matrix is
\be
\label{eq:corrmatrix}
\rho_{ij}  =  \left(
\begin{array}{ccc}
1 & 0.866 & -0.588 \\
0.866 & 1 & -0.392 \\
-0.588 & -0.392 & 1 
\end{array}
\right) \ \ \ .
\ee

\FIGURE[]{\label{fig:oblique}
\epsfig{file=FIGURES/oblique.eps,width=15cm}
\caption{$\Delta O$, for $O=T,S,U$, and $\Delta\chi^2$ as a function of the SU(2)-like Higgs mass $m_1$ for three values of the singlet-like Higgs mass $m_2$ and the mixing angle $\sin\theta$, respectively $(m_2=50\ {\rm GeV}, \sin\theta=0.5)$, black solid line, $(m_2=50\ {\rm GeV}, \sin\theta=0.1)$, red dashed line, and $(m_2=500\ {\rm GeV}, \sin\theta=0.5)$, green dot-dashed line.}}

The 95\% C.L. ellipsoid in the space of $\Delta O_i$ corresponds to taking $\Delta\chi^2 < 7.815$. The $\Delta \chi^2$ for the SM case of a single Higgs scalar with mass given by the direct search lower bound ($m_h=114.4$ GeV) is obtained by setting the $\Delta O_i=0$, yielding $\Delta\chi^2=3.546$. We consider models with the singlet scalar extension to be consistent with the EWPO if they generate values for the $\Delta O_i$ lying within the 95\% C.L. ellipsoid. The contributions from the extended scalar sector to $S$, $T$, and $U$ -- and, thereby, to electroweak observables -- is governed by three parameters: the two masses $m_{1,2}$ and mixing angle $\theta$. As indicated by Eq.~(\ref{eq:deltaT}), the dependence of the the oblique parameters on the masses an mixing angle is non-linear, so we would not expect a three-parameter fit to EWPO using $S$, $T$, and $U$ to be identical to a three-parameter fit using $m_{1,2}$ (or $\ln m_{1,2}$) and $\theta$. Nevertheless, we expect the 95\% C.L. ellipsoid for the oblique parameters to provide a reasonable indication of the consistency between EWPO and the extended scalar sector models.  

We show in Fig.~\ref{fig:oblique} our results for the differences $\Delta O$ between the oblique parameters $T,S$ and $U$ in the present model and their reference SM values, as well as the quantity $\Delta\chi^2$. We consider the variations of all of these quantities as a function of the SU(2)-like scalar mass, $m_1$, for three different combinations of the singlet-like mass $m_2$ and of the mixing angle $\sin\theta$. In particular, we consider two cases with large $\sin\theta=0.5$, the first one with small $m_2=50$ GeV~\footnote{We employ here this value of $m_2$, that would be excluded by the LEP 2 Higgs search results, for illustrative purposes only, to show the maximal effects on EWPO of having a light singlet Higgs.} and the second one with large $m_2=500$ GeV, and a third model with $m_2$ again set to 50 GeV, but smaller mixing angle, $\sin\theta=0.1$. 

It is interesting to compare the effects of $\Delta S$ and $\Delta T$ -- and their dependence on $m_{1,2}$ and $\theta$ --  on the overall $\Delta\chi^2$ (the effect of $\Delta U$ is small\footnote{The $U$ parameter is defined as the difference of finite differences involving ${\hat\Pi}_{11}$ and ${\hat\Pi}_{33}$ \cite{Degrassi:1993kn}. In the limit of vanishing weak mixing angle, the neutral scalar contributions to $U$ vanish, so it is not surprising that the scalar contributions to this parameter are quite small.}). In the case of $\Delta T$, each of the last two terms in Eq.~(\ref{eq:deltaT}) gives a negative contribution. Moreover, the magnitude of each term grows as the scalar mass is increased. Thus, $\Delta T$ tends to favor larger scalar masses and smaller mixing when $m_2 < m_1$. In contrast, $\Delta S$ favors lighter scalar masses and, therefore, larger mixing in the $m_2 < m_1$ regime. In the competition between the two effects, $\Delta S$ tends to dominate, driving the $\Delta\chi^2$ toward lighter scalars with larger mixing.

It is already known that EWPO favor a light SM Higgs, with the minimum $\chi^2$ occurring for $m_h$ below the direct search bound of $114.4$ GeV. In the case of the doublet-singlet scalar sector considered here, one would expect the precision data to favor the presence of a second mass eigenstate that is light, and that has relatively large couplings to the SM gauge bosons. This trend, indeed, emerges from our numerical analysis, as indicated in Fig.~\ref{fig:cntr}, where we show the contours of constant $\Delta\chi^2$ in the ($m_1$, $m_2$) plane for two different values of $\sin\theta$. For small $\sin\theta$ (left panel), the effect of the singlet-like scalar is weak, with the $\Delta\chi^2$ being governed almost entirely by the mass of the SU(2)-like scalar. Including a very light singlet-like scalar improves the $\Delta\chi^2$, but not substantially. In contrast, for large $\sin\theta$, corresponding to mass eigenstates that are nearly equal mixtures of singlet and SU(2)$_L$ components, one can improve the $\Delta\chi^2$ from the SM reference value if at least one of the two scalars is light. The LEP Higgs searches still exclude much of this region, since the mixing is nearly maximal. Nevertheless, the EWPO allow for both scalars to be considerably heavier than $114.4$ GeV (up to about 220 GeV) in the presence of maximal mixing. 

From these considerations, we expect that the impact of EWPO on models featuring a strongly first order phase transition will be to limit $m_1$ to values smaller than around 200 GeV; large values of the mass of the singlet-like scalar are EWPO-viable as long as the mixing with the SU(2)-like scalar is suppressed. We  confirm this trend comparing the upper and lower panels of Fig.~\ref{fig:m1m2}, where we show the same models we showed in the previous plots in the $(m_1,m_2)$ plane, without (upper panels) and with (lower panels) the EWPO 95\% C.L. constraint. Imposing the latter eliminates nearly all models with a very light singlet-like scalar and heavy ($m_1\gtrsim 200$ GeV) SU(2)-like scalar. 

\FIGURE[]{\label{fig:cntr}
\mbox{\epsfig{file=FIGURES/cntr1.eps,width=7.5cm}\qquad
\epsfig{file=FIGURES/cntr2.eps,width=7.5cm}}
\caption{Curves at fixed $\Delta\chi^2$ in the ($m_1,m_2$) plane, for two values of the mixing angle $\sin\theta=0.2$ (left panel) and 0.7 (right panel).}}

\vskip 0.1in
\FIGURE[]{
\label{fig:m1m2}
\epsfig{file=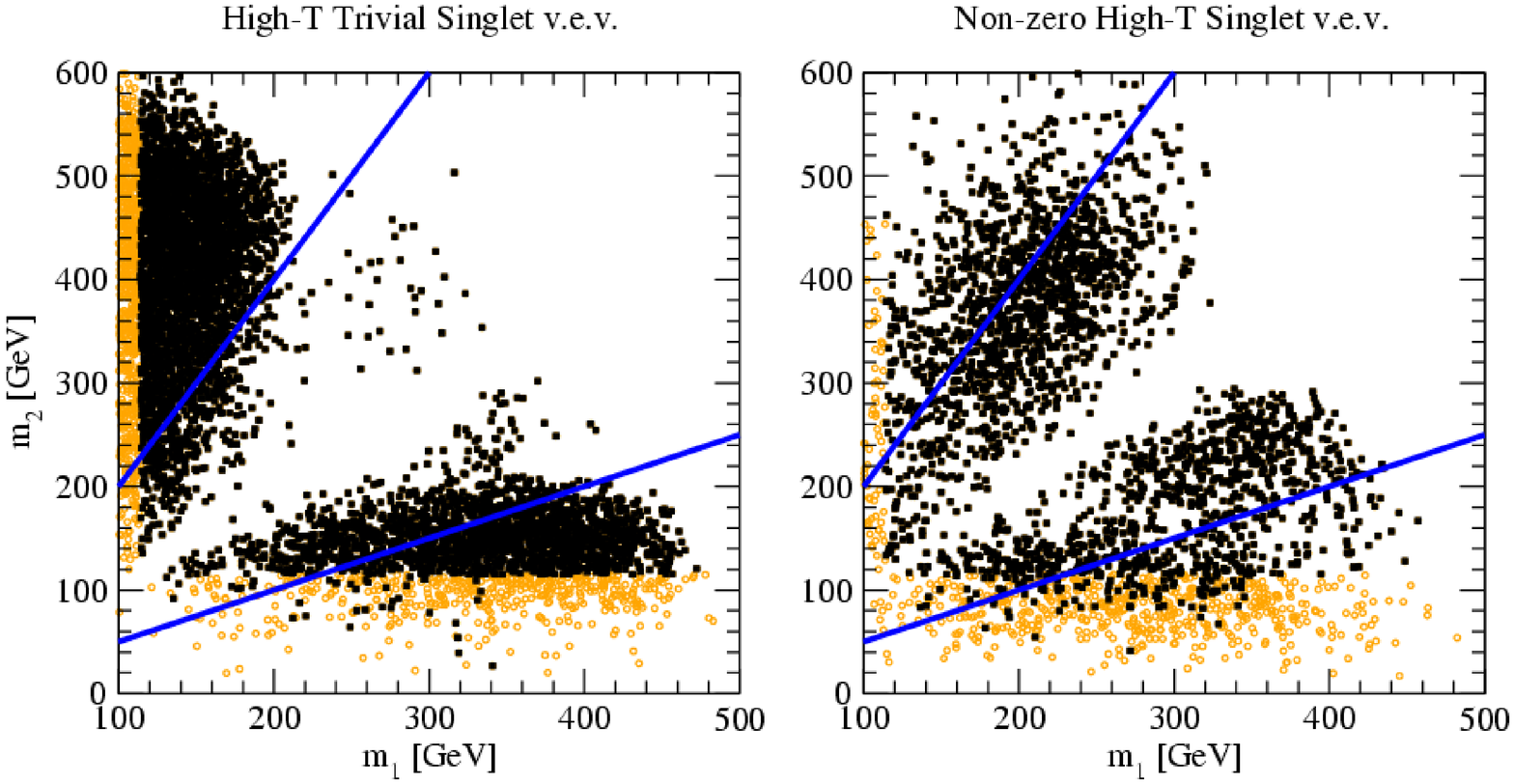,width=15cm}\\
\epsfig{file=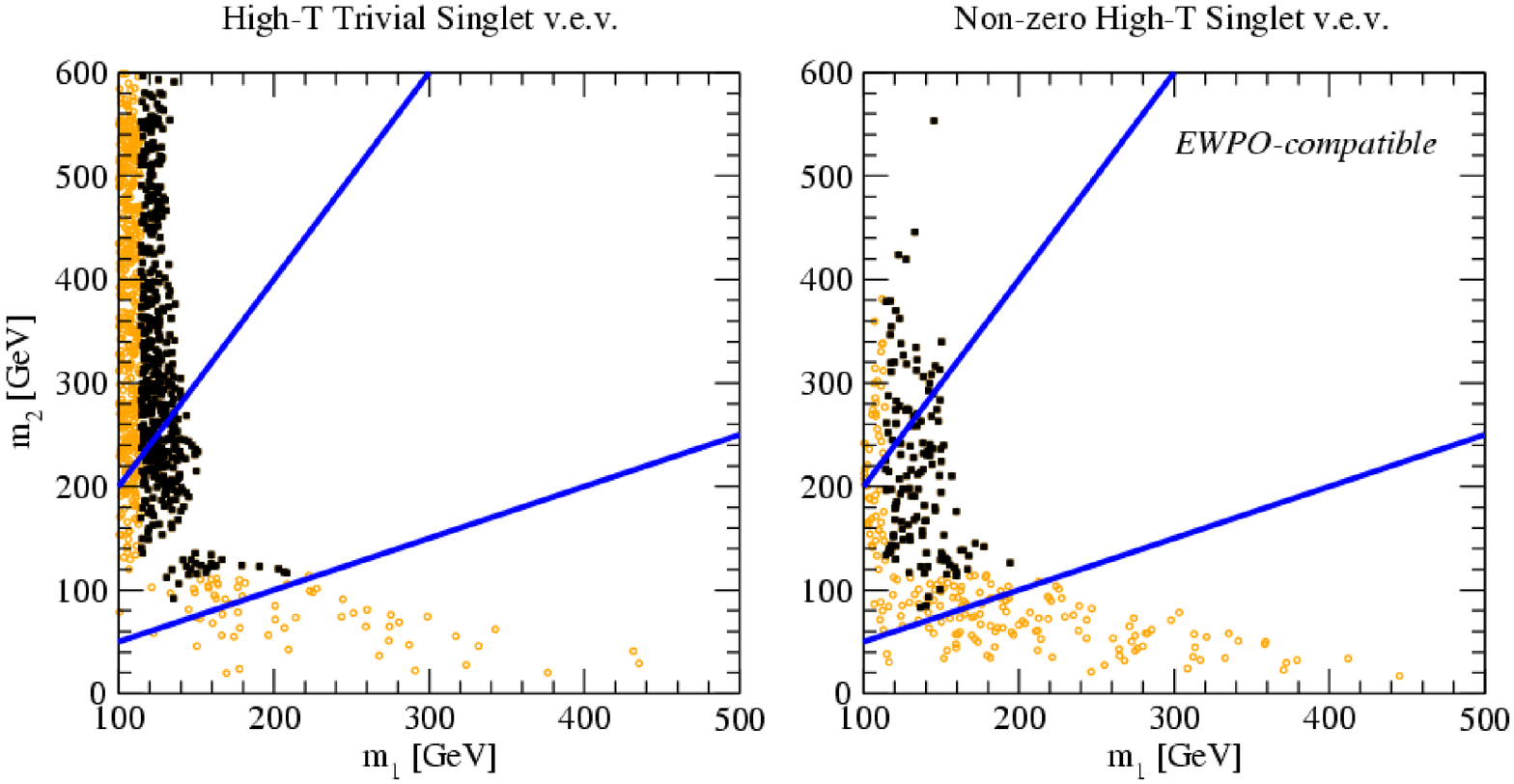,width=15cm}
\caption{As in Fig.~\ref{fig:critalpha}, for the physical Higgs masses ($m_1$ corresponds to the mostly SU(2)-like mass eigenstate, and $m_2$ to the mostly singlet-like mass eigenstate). The two upper panels show all models, while the lower panels show the models that are not ruled out at 95\% C.L. by EWPO. For models in the lower-right part of the plot (below the lower blue line), the decay $h_1\rightarrow h_2\ h_2$ is open, while for models above the upper blue line the decay $h_2\rightarrow h_1\ h_1$ is open.}}

\noindent{\em Collider Implications}
\vskip 0.1in

As a starting point of our analysis of collider implications, we refer to Eqs.~(\ref{eq:massparams}-\ref{eq:shiftparams}) that indicate how the masses and mixing of the fields characterizing fluctuations about the zero temperature vevs depend on the parameters in the full potential. We first consider the possibility that Higgs production and decays at the LHC could be modified by the singlet extension as recently discussed in Ref.~\cite{O'Connell:2006wi} and elsewhere \cite{singletcollideral}. If $m_1\geq 2m_2$, then the decay $h_1\to h_2h_2$ becomes kinematically allowed, and the branching ratio (BR) of $h_1$ to the conventional SM final states will be reduced. The presence of this additional Higgs decay channel can provide interesting signals as four $b$-jets or $2\tau + 2 b$; this possibility has been explored in various supersymmetric extensions \cite{hto4sm}. Conversely, if $m_2\geq 2m_1$, then double SM-like Higgs production becomes possible if $h_2$ is first produced, leading to similar exotic final states. However, the production rate of $h_2$ is suppressed relative to $h_1$ by the mixing factor $\tan^2\theta < 1$.

Fig.~\ref{fig:m1m2} illustrates the distribution of points in the $(m_1,m_2)$ plane that give rise to a strong first order EWPT, with (lower panels) and without (upper panels) imposing the EWPO constraint. As before, the distribution of all models is given by the light and dark points, while those consistent with LEP searches are only those in black. Notice that we find several models featuring a large value of the singlet-like Higgs mass eigenstate $m_2$: for those models, since $m_2\gg T_c$, finite temperature corrections play a negligible role in making the EWPT strongly first order. Rather, the driving effect in the Higgs effective potential is played by tree level terms involving the field $S$, as discussed in Sec.~\ref{sec:ewpt}. Points lying below the $m_2=m_1/2$ blue line correspond to kinematically allowed $h_1\to h_2h_2$ decays, while those above the blue $m_2=2m_1$ line indicate models in which the $h_2\to h_1 h_1$ channel is open. 

If one does not impose the EWPO constraints,  a substantial fraction of the EWPT-viable models also give rise to either of these two open channels, indicating the possibility that such models may be probed with Higgs studies at the LHC. A detailed analysis of such searches will appear in a companion paper \cite{higgssearchinprep}. It is striking, however,  that imposing the EWPO constraints closes -- for a large fraction of the models in our scan -- the window of $h_1\to h_2h_2$ decays for models also compatible with LEP\footnote{Models in this region still occur, but they do not appear in the plot due to the resolution of our scan. Their existence is more apparent in the special case of $\mathbb{Z}_2$ symmetry discussed below.}. In contrast, the decay $h_2\to h_1h_1$ is generically open for a large set of models, provided  that $m_1\lesssim160$ GeV.

\FIGURE[!t]{
\mbox{\hspace*{-1.5cm}\epsfig{file=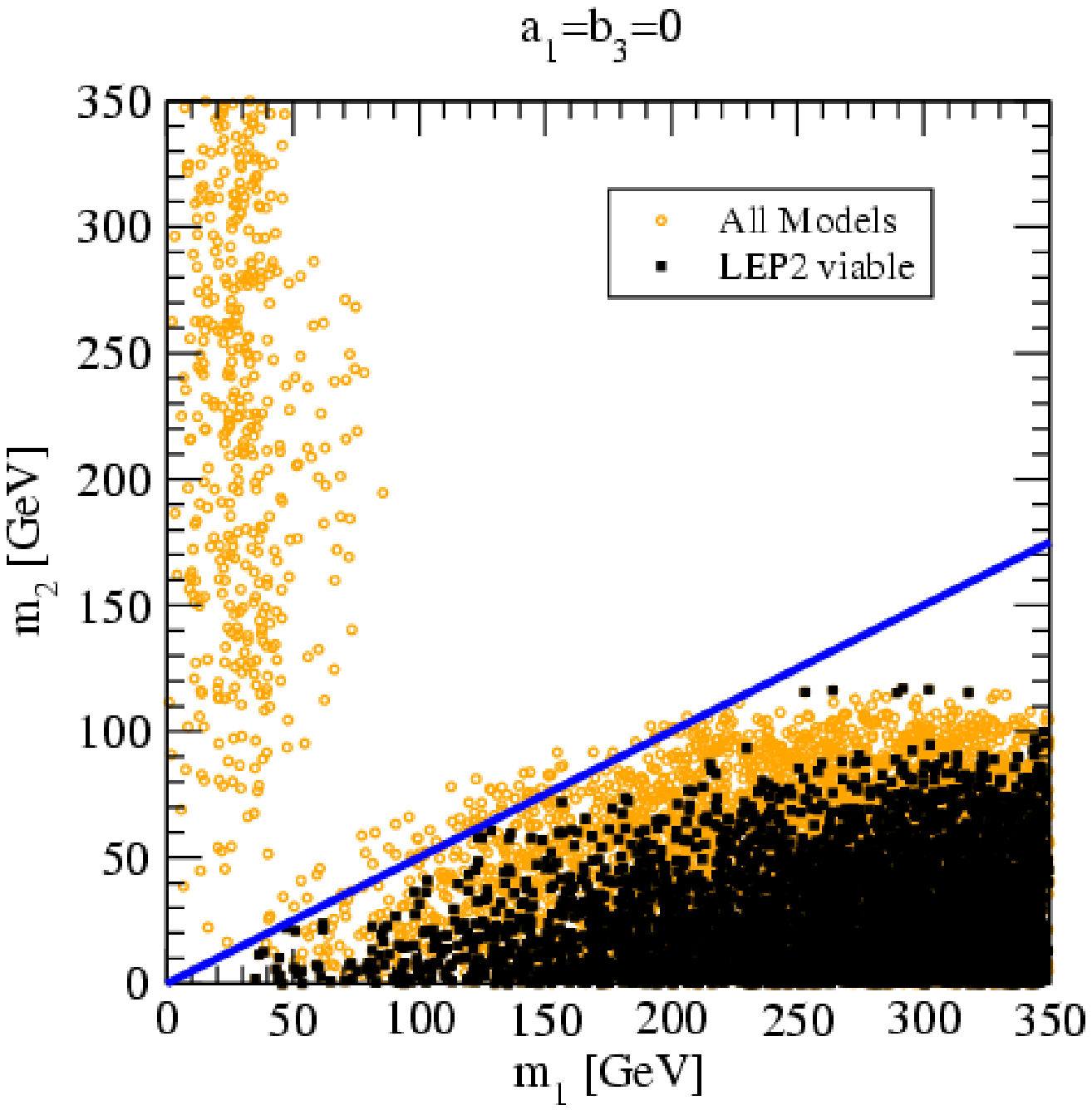,width=7.5cm}\qquad
\epsfig{file=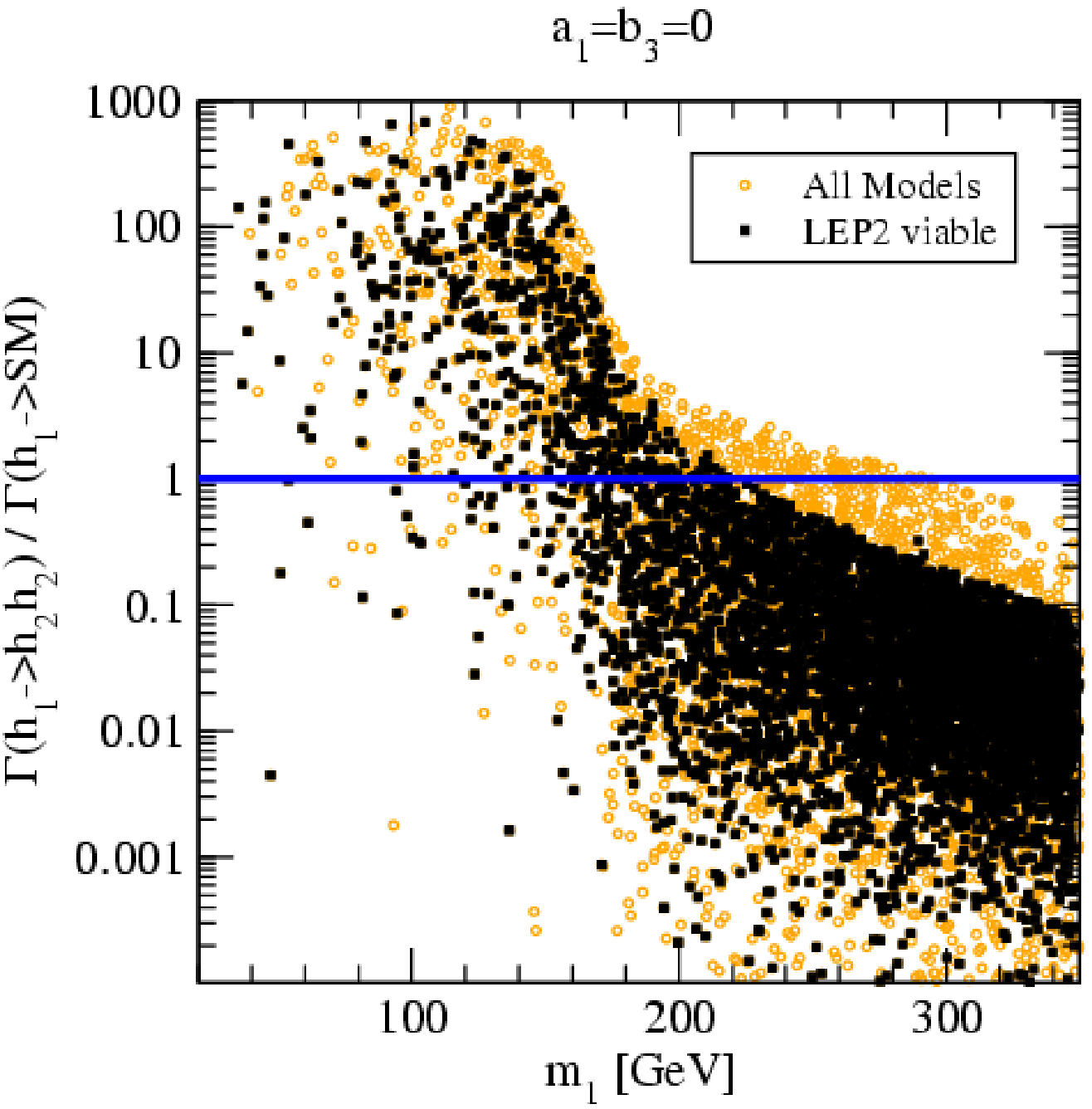,width=7.5cm}}\\
\mbox{\hspace*{-1.5cm}\epsfig{file=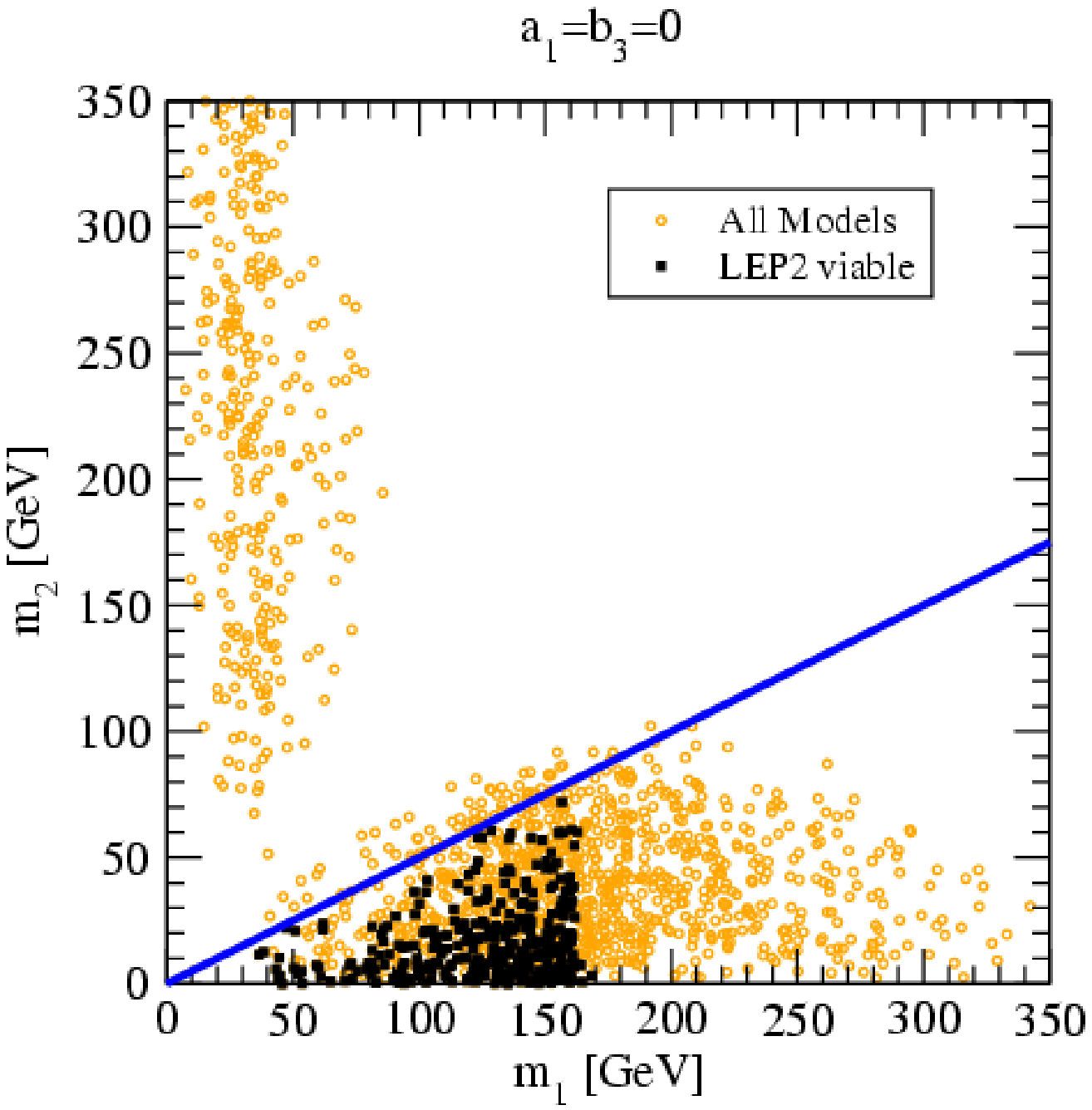,width=7.5cm}\qquad
\epsfig{file=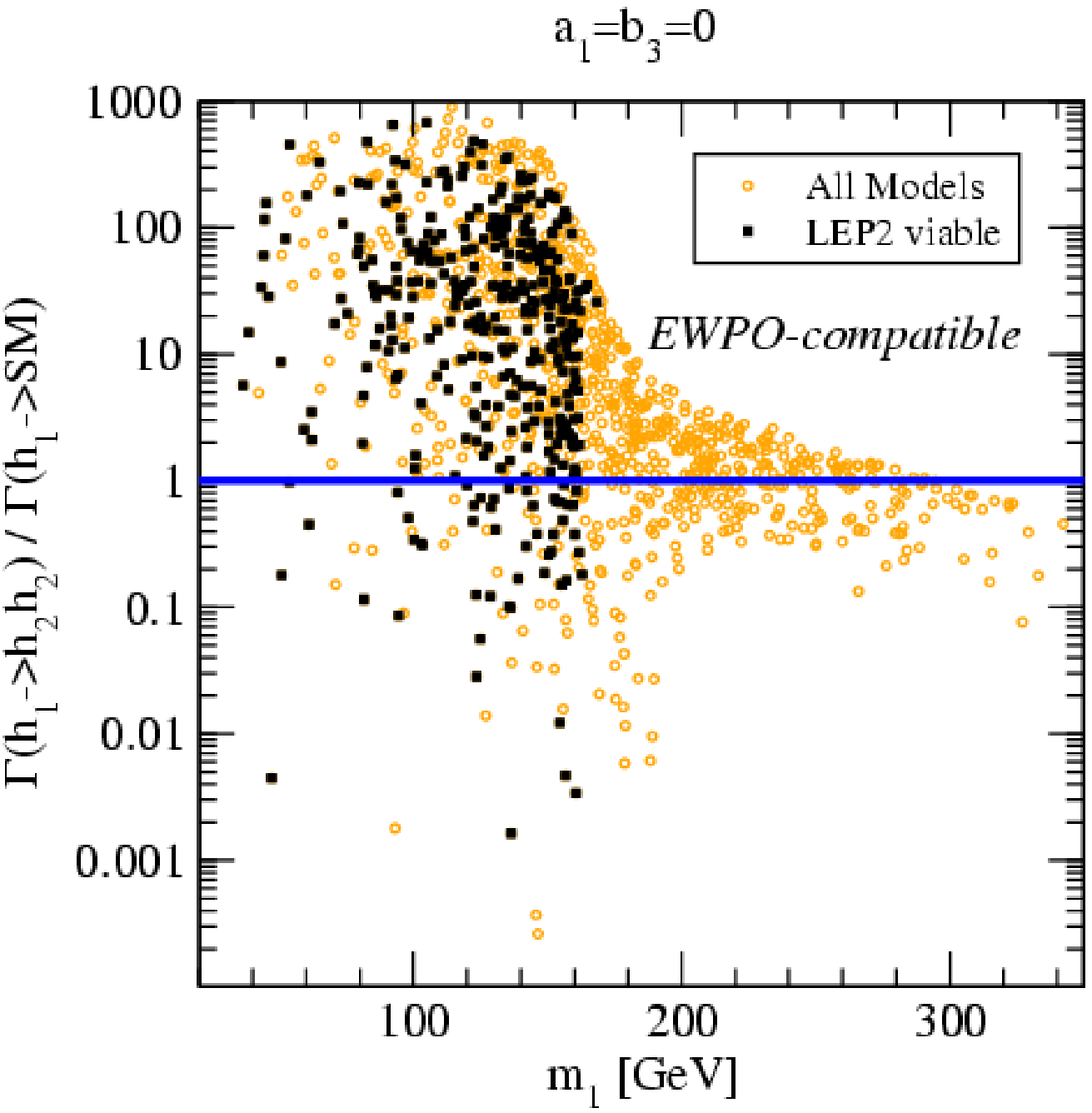,width=7.5cm}}\\

\caption{Right: The $(m_1,m_2)$ plane for models featuring $a_1=b_3=0$. Left: for the same models, the ratio of the decay width of the SU(2)-like Higgs into two singlet, over the total decay width of the SU(2)-like Higgs into Standard Model particles. The two upper panels show all models, while the lower panels show the models that are not ruled out at 95\% C.L. by EWPO.\label{fig:sms1}}}
As a special case, we focus on models having a $\mathbb{Z}_2$ symmetry before SSB, given by the condition $a_1=b_3=0$. Such models are particularly interesting since they also admit the possibility that $h_2$ is the dark matter particle, since it cannot decay to any SM particles when $x_0=0$ and there is no mixing with the SM-like Higgs \cite{Silveira:1985rk}. The corresponding distribution of EWPT-viable models in the $(m_1,m_2)$ plane is given in the left panels of Fig.~\ref{fig:sms1}. Notice that here we consider a special slice of the parameter space, as we scan only over four of the six parameters in Tab~\ref{tab:param}, setting $a_1=b_3=0$. Thus,  the ensemble of $\sim 10^4$ models we find differs from those we considered above. 

An important feature emerging from this scan is that the LEP-consistent models always lie in the $h_1\to h_2 h_2$ allowed region. Even after EWPO constraints are imposed, a large set of models exist for which this decay channel is open. For these models, one would expect a reduction in the BR of $h_1$ to SM modes. The degree of this reduction depends on the rate 
\be
\label{eq:rate}
\Gamma(h_1\to h_2 h_2) = \frac{g^2_{122}}{8\pi m_1}\, \sqrt{1-4m_2^2/m_1^2}\ \ \ ,
\ee
where
\begin{eqnarray}\label{eq:g122}
\nonumber g_{122}&=&\frac{v_0\ c_\theta}{2}\left(a_2(c^2_\theta-2s_\theta^2)+6\lambda s^2_\theta\right)+\frac{a_1+2a_2x_0}{2}\left(s_\theta^3-c^2_\theta s_\theta\right)+\\
&&+(b_3+3b_4x_0)\ c^2_\theta s_\theta,\qquad\quad [s_\theta\equiv\sin\theta,\ c_\theta\equiv\cos\theta].
\end{eqnarray}
We note in passing that if $\Gamma(h_1\to h_2 h_2)\gtrsim\Gamma^{\rm SM}$, models with a light $SU(2)$-like Higgs featuring $m_1<114.4$ GeV can be compatible with the LEP exclusion limits \cite{Barate:2003sz} [see Eq.~(\ref{eq:xi})]. The rate for decay to SM particles $\Gamma(h_1\to\, {\rm SM\, modes})$ depends on the final states that dominate for a given value of $m_1$. For light $h_1$, its SM decays are dominated by the $b{\bar b}$ final state. In Ref.~\cite{O'Connell:2006wi}, the BR reduction was studied for this regime under the implicit assumption that $x_0=0$ ($\sin\theta=0$). In this case, $\mu_s^2=b_2/2+a_2v_0^2/2$ with $b_2$ an independent parameter, so that increasing the coupling $a_2$ for fixed $b_2$ affects the rate in Eq.~(\ref{eq:rate}) through both the overall coupling strength as well as the mass of the $h_2$.

Here we consider the more general situation (compared to the foregoing) in which $x_0\not=0$ is an independent parameter. In the right panels of Fig.~\ref{fig:sms1} we show the ratio $\Gamma(h_1\to h_2 h_2)/\Gamma(h_1\to\, {\rm SM\, modes}$) for the EWPT-viable models, before (upper panel) and after (lower panel) imposing EWPO constraints. For light $h_1$ ($114\lesssim m_1\lesssim 180$ GeV), this ratio can be quite large, leading to a suppression of the SM BR by an order of magnitude or more. For heavier $h_1$, the decay competes with decays to massive gauge bosons, leading to a reduction in the SM BR from one to more than ten percent, while for $m_1\gtrsim 300$ GeV, the reduction is generally less than ten percent and for many models can be well below one percent. However, as can be seen in the lower right panel, these models are  incompatible with EWPO constraints. For models consistent with EWPO at the 95\% C.L. we always expect a BR suppression at at least the percent level. In short, if a SM-like Higgs with $m_1\lesssim 200$ GeV is discovered at the LHC, one would expect an observable reduction in its SM BR in the presence of a light singlet-like scalar that gives rise to a strong first order EWPT. For those models giving a BR reduction of $\lesssim 10\%$, Higgs studies at a Linear Collider would likely be needed to observe this suppression. Conversely, the absence of a BR reduction for heavier SM-like Higgses would point to models that do not have a $\mathbb{Z}_2$ symmetry before SSB or have a heavier singlet-like scalar. In this case,  one would look for signatures either through exotic final states at the LHC, as discussed above, or through future Higgsstrahlung studies at a linear collider.

\FIGURE[]{\label{fig:m1m2ilc}
\mbox{\epsfig{file=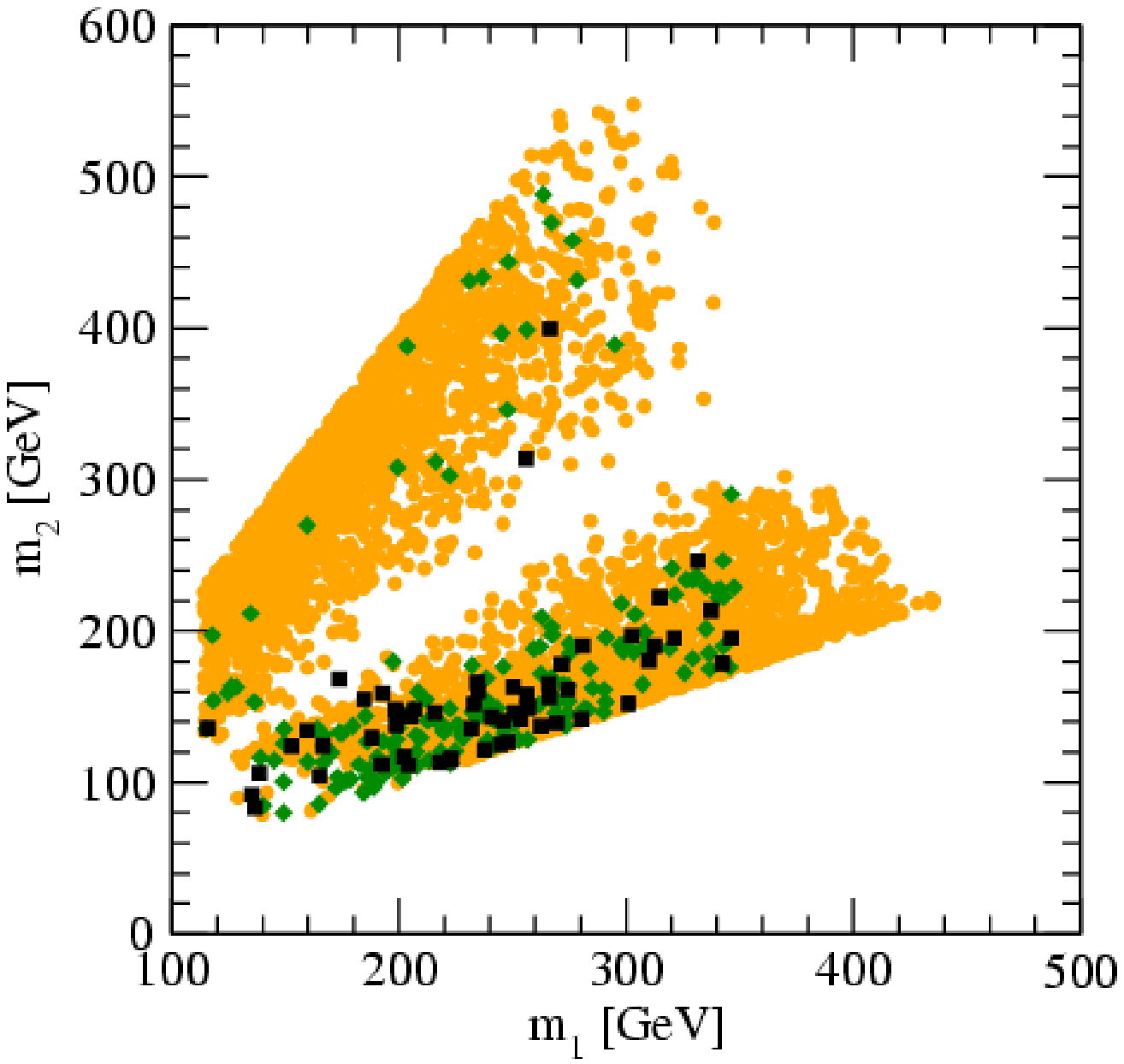,width=7.5cm}\qquad\epsfig{file=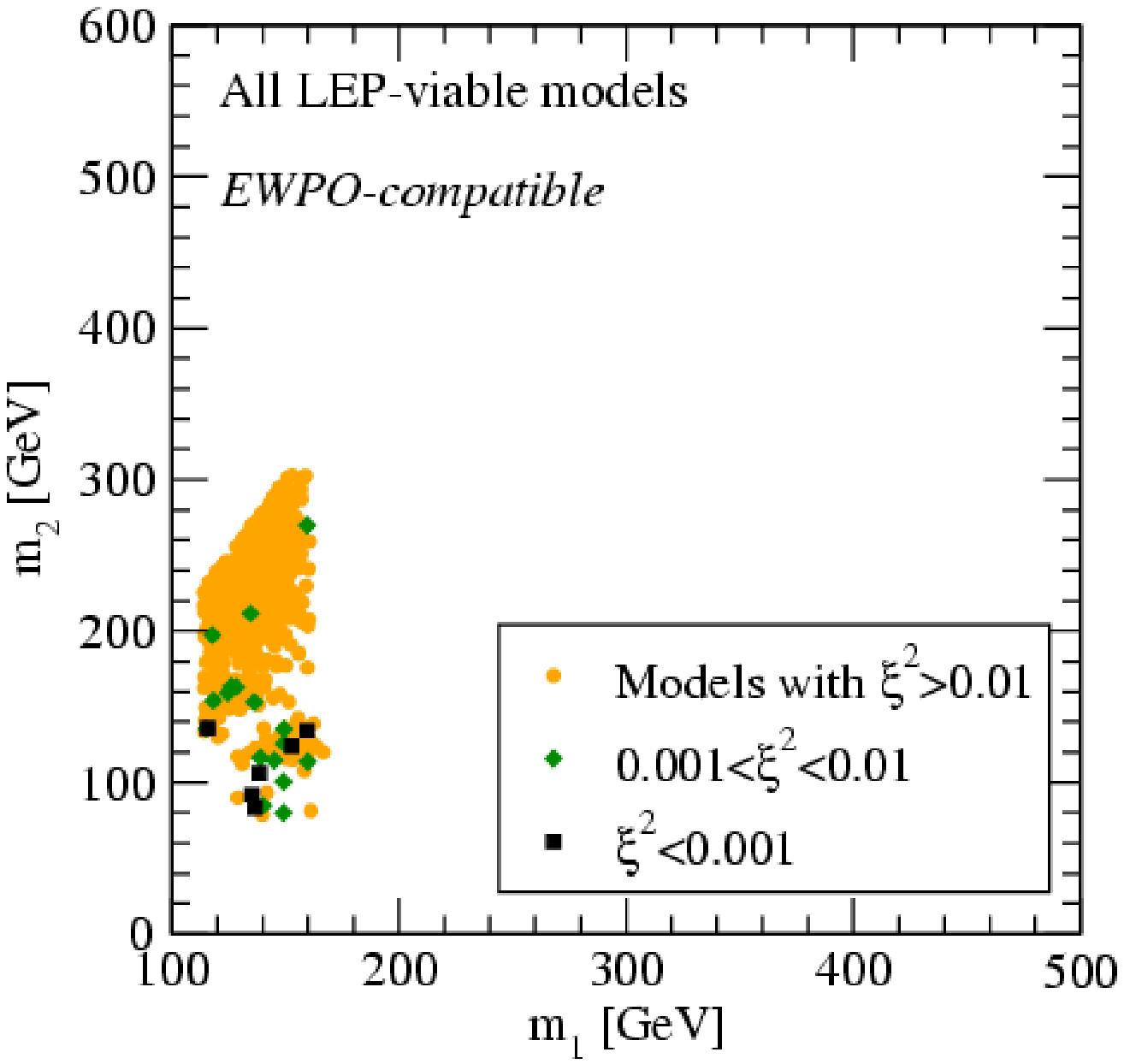,width=7.5cm}}
\caption{The $(m_1,m_2)$ plane for all models considered in the various scans, in the region where the decays $h_{1,2}\rightarrow h_{2,1}h_{2,1}$ are kinematically forbidden. In the right panel we removed models that did not satisfy the EWPO constraints at 95\% C.L.}}
In scenarios having no $\mathbb{Z}_2$ symmetry, there exists a substantial fraction of EWPT-viable models that lie between the two blue lines in Fig.~\ref{fig:m1m2} and that imply no change in the Higgs decay modes at the LHC. For these models, it is interesting to ask whether future Higgsstrahlung studies at a Linear Collider could provide a useful probe. To that end, we show in Fig.~\ref{fig:m1m2ilc} the distribution of points where $h_{1,2}\to h_{2,1} h_{2,1}$ decays are kinematically forbidden, focusing on those having a small value for $\xi^2$ that characterizes the reduction in the Higgsstrahlung rate. Points with $m_2< 114.4$ GeV that are not ruled out by the LEP Higgs search generally have $\xi^2 << 1$. For $m_{1,2} > 114.4$ GeV, the overwhelming majority of models have $\xi^2>0.01$ and generally correspond to large values of $a_1$ as implied by Eqs.~(\ref{eq:massparams},\ref{eq:mixing}). Those points are shown with orange dots in the figure. Points with $0.001< \xi^2 < 0.01$ (dark green diamonds) and $\xi^2< 0.001$ (black squares) indicate the Higgsstrahlung sensitivity needed to probe these models. To the extent that Higgsstrahlung studies at a future $e^+e^-$ collider can search for neutral scalars in this mass range with Higgsstrahlung rates up to 100 times smaller than for a SM-like Higgs, one could probe a large class of EWPT-viable singlet models that would not be readily accessible at the LHC. Finally, the panel to the right illustrates the impact of imposing EWPO constraints on the models at hand. As elsewhere, including these constraints considerably reduces the space of viable models. 

\section{Conclusions}
\label{sec:conclusions}

Searches for the Higgs boson at the LHC and studies of its properties both there and at a future Linear Collider will be an important thrust at these colliders. In this study, we have analyzed the implications for Higgs boson phenomenology and the EWPT of the simplest extension of the Standard Model scalar sector containing one additional gauge singlet scalar field. Our goal has been to identify general features of more complicated models with gauge singlet scalars, such as extensions of the MSSM in which the $\mu$-term is generated dynamically as the vev of such a scalar field. Although some details of Higgs phenomenology as it relates to the EWPT will depend on specific model realizations, we believe we have outlined general trends that one would expect in most scenarios of this type. In doing so,  we derived a simple criterion that allows one to predict, starting from the parameters in the scalar potential, whether or not the singlet field acquires a vev prior to the EWPT. We also obtained a compact expression [Eq.~(\ref{eq:crit2})] for the condition of obtaining a strong first order EWPT as needed to prevent washout of the BAU by electroweak sphalerons, under approximations that were systematically cross-checked through our numerical analysis.

The features that emerge from our study include:

\begin{itemize}
\item[-] We find that a strongly first order EWPT can occur quite readily in models where the singlet scalar obtains a non-zero vev ($x_0\ne0$). Choosing $x_0=0$ (see e.g. \cite{Espinosa:2007qk}) is a necessary condition for models that have a scalar dark matter candidate and forces the potential to have a $\mathbb{Z}_2$ symmetry. Conversely, models with a $\mathbb{Z}_2$ symmetry before SSB do not necessarily have $x_0=0$.
\item[-] Models with singlet scalars can readily enhance the strength of the EWPT to allow viable electroweak baryogenesis through one of three tree-level effects: the presence of a sufficiently large negative coupling between $(H^\dag H)$ and $S$, thereby enhancing the numerator of Eq.~(\ref{eq:crit2}); the presence of a sufficiently  large negative coupling between $(H^\dag H)$ and $S^2$, thereby reducing the denominator of Eq.~(\ref{eq:crit2}); or the choice of parameters in the singlet scalar potential that generates a non-trivial high temperature minimum (before electroweak symmetry is broken). 
\item[-] The presence of a $\mathbb{Z}_2$-violating  $(H^\dag H)S$ coupling leads to mixing between the singlet and neutral SU(2)$_L$ scalars, as does the $\mathbb{Z}_2$-invariant $(H^\dag H)S^2$ interaction when $x_0\neq 0$. Higgs boson searches at LEP exclude models in which one or the other of these couplings is large enough to induce a strong first order EWPT when one of the scalar masses is well below the direct search lower bound. Future Higgsstrahlung studies at a Linear Collider could provide an additional probe of these effects for heavier scalars\footnote{In principle, studies of Higgs production via $WW$ and $ZZ$ fusion could also be used for this purpose.}.
\item[-] Considerations of EWPO imply that EWPT-viable models in which the singlet-like scalar can decay to two SU(2)-like scalars occur more copiously than those in which the SU(2)-like scalar decays to a pair of singlet-like scalars. In principle, the models in which the singlet-like scalar decays to a pair of SU(2)-like scalars could be identified through the presence of exotic final states at the LHC, such as four $b$-jets, $b \bar b \tau^+ \tau^-$, or $b{\bar b}\gamma\gamma$. EWPT-viable models in which these decays are kinematically forbidden could still be tested with Higgsstrahlung studies at a Linear Collider. 
\item[-] In the special case of models having $\mathbb{Z}_2$ symmetry before SSB, all EWPT-viable models that are consistent with EWPO lie in the region where the SU(2)-like scalar can decay to pairs of the singlet-like scalars. For these models, one would expect to see large suppressions of the branching ratio to SM Higgs decay modes, as the branching ratio to the pair of singlet-like scalars would be considerable. 
\end{itemize}

In short, the possibility of testing the viability of producing the strong first order EWPT needed for successful electroweak baryogenesis in models with a small number of singlet scalars appears to be feasible using a combination of Higgs studies at the LHC and a future Linear Collider. Together with the next generation of  \lq\lq table top" searches for the electric dipole moments of the electron, neutron, neutral atoms, and nuclei that will provide powerful probes of new electroweak CP-violation, up-coming collider Higgs studies could help unravel the origin of baryonic matter. 

\acknowledgments

We thank M. Wise, P. Langacker, D. Chung, V. Barger and A. Kusenko for several helpful discussions, and D. O'Connell for useful conversations and a critical reading of the manuscript. We thank J. Kile and J. Erler for assistance with the GAPP routine. This work  was supported by the U.S. Department of Energy contracts DE-FG02-05ER41361(M.R.-M. and S.P), DE-FG03-92-ER40701 (S.P.) and DE-FG02-95ER40896 (G.S.); by the National Science Foundation grant PHY-0555674 (M.R.-M.) and by NASA grant number NNG05GF69G (S.P.).
	
%


\begin{thebibliography}{99}

\bibitem{Yao:2006px}
  W.~M.~Yao {\it et al.}  [Particle Data Group],
  J.\ Phys.\ G {\bf 33} (2006) 1.

\bibitem{Spergel:2006hy}
  D.~N.~Spergel {\it et al.}  [WMAP Collaboration],
  arXiv:astro-ph/0603449.

\bibitem{Sakharov:1967dj}
A.~D.~Sakharov,
Pisma Zh.\ Eksp.\ Teor.\ Fiz.\  {\bf 5}, 32 (1967)
[JETP Lett.\  {\bf 5}, 24 (1967)].

\bibitem{Jarlskog:1985ht}
  C.~Jarlskog,
  Phys.\ Rev.\ Lett.\  {\bf 55}, 1039 (1985).

\bibitem{Shaposhnikov:1987tw}
M.~E.~Shaposhnikov,
Nucl.\ Phys.\ B {\bf 287}, 757 (1987).

\bibitem{Riotto:1999yt}
A.~Riotto and M.~Trodden,  
Ann.\ Rev.\ Nucl.\ Part.\ Sci.\  {\bf 49}, 35 (1999)  [arXiv:hep-ph/9901362].

\bibitem{Dine:2003ax}
  M.~Dine and A.~Kusenko,
  Rev.\ Mod.\ Phys.\  {\bf 76}, 1 (2004)
  [arXiv:hep-ph/0303065].

\bibitem{Pospelov:2005pr} M.~Pospelov and A.~Ritz,
Annals Phys.\  {\bf 318}, 119 (2005)  [arXiv:hep-ph/0504231].

\bibitem{Erler:2004cx}
  J.~Erler and M.~J.~Ramsey-Musolf,
  Prog.\ Part.\ Nucl.\ Phys.\  {\bf 54}, 351 (2005)
  [arXiv:hep-ph/0404291].

\bibitem{Ramsey-Musolf:2006vr}
  M.~J.~Ramsey-Musolf and S.~Su,
  arXiv:hep-ph/0612057.

\bibitem{Rummukainen:1998as}
  K.~Rummukainen, M.~Tsypin, K.~Kajantie, M.~Laine and M.~E.~Shaposhnikov,
  Nucl.\ Phys.\  B {\bf 532} (1998) 283
  [arXiv:hep-lat/9805013].

\bibitem{Barate:2003sz}
  R.~Barate {\it et al.}  [LEP Working Group for Higgs boson searches],
  Phys.\ Lett.\  B {\bf 565} (2003) 61
  [arXiv:hep-ex/0306033].

\bibitem{ewbstop}

M.~Carena, M.~Quiros, A.~Riotto, I.~Vilja and C.~E.~Wagner,
Nucl.\ Phys.\ B {\bf 503}, 387 (1997) [arXiv:hep-ph/9702409];\\
  M.~Carena, M.~Quiros and C.~E.~M.~Wagner,
  Nucl.\ Phys.\ B {\bf 524}, 3 (1998)
  [arXiv:hep-ph/9710401];\\
J.~M.~Cline, M.~Joyce and K.~Kainulainen,
JHEP {\bf 0007}, 018 (2000) [arXiv:hep-ph/0006119]. Erratum:
arXiv:hep-ph/0110031;\\
M.~Carena, J.~M.~Moreno, M.~Quiros, M.~Seco and C.~E.~Wagner,
Nucl.\ Phys.\ B {\bf 599}, 158 (2001) [arXiv:hep-ph/0011055];\\
  M.~Carena, M.~Quiros, M.~Seco and C.~E.~M.~Wagner,
  Nucl.\ Phys.\ B {\bf 650} (2003) 24
  [arXiv:hep-ph/0208043].

\bibitem{MSSMHiggs}LEP Working Group for Higgs Boson Searches, Search for
Neutral MSSM Higgs Bosons at LEP, Report No. LHWG-Note 2005-01.

\bibitem{Pietroni:1992in}
  M.~Pietroni,
  Nucl.\ Phys.\  B {\bf 402} (1993) 27
  [arXiv:hep-ph/9207227].

\bibitem{othersinglet} See {\em e.g.}
  A.~T.~Davies, C.~D.~Froggatt and R.~G.~Moorhouse,
  Phys.\ Lett.\  B {\bf 372} (1996) 88
  [arXiv:hep-ph/9603388];\\
  S.~J.~Huber and M.~G.~Schmidt,
singlet,''
  Nucl.\ Phys.\ B {\bf 606} (2001) 183
  [arXiv:hep-ph/0003122];\\
  A.~Menon, D.~E.~Morrissey and C.~E.~M.~Wagner,
  Phys.\ Rev.\ D {\bf 70}, 035005 (2004);\\
  S.~W.~Ham, S.~K.~OH, C.~M.~Kim, E.~J.~Yoo and D.~Son,
  Phys.\ Rev.\  D {\bf 70}, 075001 (2004),
  [arXiv:hep-ph/0406062];\\
  K.~Funakubo, S.~Tao and F.~Toyoda,
  Prog.\ Theor.\ Phys.\  {\bf 114} (2005) 369
  [arXiv:hep-ph/0501052];\\
  S.~J.~Huber, T.~Konstandin, T.~Prokopec and M.~G.~Schmidt,
  Nucl.\ Phys.\  B {\bf 757} (2006) 172
  [arXiv:hep-ph/0606298];\\
  C.~Balazs, M.~Carena, A.~Freitas and C.~E.~M.~Wagner,
  arXiv:0705.0431 [hep-ph].

\bibitem{langauprime}
For the case of models with a $U(1)^\prime$ see {\em e.g.}  J.~Kang, P.~Langacker, T.~j.~Li and T.~Liu,
  Phys.\ Rev.\ Lett.\  {\bf 94} (2005) 061801
  [arXiv:hep-ph/0402086] and references therein.

\bibitem{muproblem}
  J.~R.~Ellis, J.~F.~Gunion, H.~E.~Haber, L.~Roszkowski and F.~Zwirner,
  Phys.\ Rev.\  D {\bf 39} (1989) 844. For a recent overview and references see
also: V.~Barger, P.~Langacker, H.~S.~Lee and G.~Shaughnessy,
  Phys.\ Rev.\  D {\bf 73} (2006) 115010
  [arXiv:hep-ph/0603247].

\bibitem{singletkusenko}
  See   A.~Kusenko,
  Phys.\ Rev.\ Lett.\  {\bf 97} (2006) 241301
  [arXiv:hep-ph/0609081] and references therein.

\bibitem{O'Connell:2006wi}
  D.~O'Connell, M.~J.~Ramsey-Musolf and M.~B.~Wise,
  arXiv:hep-ph/0611014.


\bibitem{Espinosa:2007qk}
J.~R.~Espinosa and M.~Quiros,
arXiv:hep-ph/0701145.

\bibitem{wjm06} W. J. Marciano, talk given at Jefferson Lab Electroweak
Workshop, Newport News, VA (December, 2006).

\bibitem{Anderson:1991zb}
  G.~W.~Anderson and L.~J.~Hall,
  Phys.\ Rev.\  D {\bf 45}, 2685 (1992).


\bibitem{extrasinglet} See e.g.
  J.~Choi and R.~R.~Volkas,
  Phys.\ Lett.\  B {\bf 317} (1993) 385
  [arXiv:hep-ph/9308234];\\
  K.~E.~C.~Benson,
  Phys.\ Rev.\  D {\bf 48} (1993) 2456;\\
  J.~R.~Espinosa and M.~Quiros,
  Phys.\ Lett.\  B {\bf 305} (1993) 98
  [arXiv:hep-ph/9301285];\\
    J.~Choi,
  Phys.\ Lett.\  B {\bf 345} (1995) 253
  [arXiv:hep-ph/9409360];\\
  L.~Vergara,
  Phys.\ Rev.\  D {\bf 55} (1997) 5248;\\
  S.~W.~Ham, Y.~S.~Jeong and S.~K.~Oh,
  J.\ Phys.\ G {\bf 31} (2005) 857
  [arXiv:hep-ph/0411352].

\bibitem{Silveira:1985rk}
  V.~Silveira and A.~Zee,
  Phys.\ Lett.\  B {\bf 161} (1985) 136. See also H.~Davoudiasl, R.~Kitano,
T.~Li and H.~Murayama,
  Phys.\ Lett.\  B {\bf 609} (2005) 117
  [arXiv:hep-ph/0405097] and references therein.

\bibitem{Coleman:1973jx}
  S.~R.~Coleman and E.~Weinberg,
  Phys.\ Rev.\  D {\bf 7}, 1888 (1973).

\bibitem{Dolan:1973qd}
  L.~Dolan and R.~Jackiw,
  Phys.\ Rev.\  D {\bf 9}, 3320 (1974).

\bibitem{Ahriche:2007jp}
  A.~Ahriche,
  arXiv:hep-ph/0701192.

\bibitem{espinosa}
  P.~Arnold and O.~Espinosa,
  Phys.\ Rev.\  D {\bf 47} (1993) 3546
  [Erratum-ibid.\  D {\bf 50} (1994) 6662]
  [arXiv:hep-ph/9212235].


\bibitem{Quiros:1999jp}
  M.~Quiros,
  arXiv:hep-ph/9901312.

\bibitem{Ham:2004cf}  See e.g. 
  K.~Enqvist, K.~Kainulainen and I.~Vilja,
  Nucl.\ Phys.\  B {\bf 403} (1993) 749;\\  
  N.~Sei, I.~Umemura and K.~Yamamoto,
  Phys.\ Lett.\  B {\bf 299} (1993) 286;\\
  S.~W.~Ham, Y.~S.~Jeong and S.~K.~Oh,
  J.\ Phys.\ G {\bf 31} (2005) 857
  [arXiv:hep-ph/0411352].

\bibitem{Wood:1997zq}
  C.~S.~Wood, S.~C.~Bennett, D.~Cho, B.~P.~Masterson, J.~L.~Roberts,
C.~E.~Tanner and C.~E.~Wieman,
  Science {\bf 275}, 1759 (1997).

\bibitem{Anthony:2003ub}
  P.~L.~Anthony {\it et al.}  [SLAC E158 Collaboration],
  Phys.\ Rev.\ Lett.\  {\bf 92}, 181602 (2004)
  [arXiv:hep-ex/0312035].

\bibitem{radiativeinprep}
  M.~J.~Ramsey-Musolf et al., in preparation

\bibitem{gapp}
  J.~Erler,
  arXiv:hep-ph/0005084.

\bibitem{unknown:2007bx}
    [CDF Collaboration],
  arXiv:hep-ex/0703034.

\bibitem{Degrassi:1993kn}
  G.~Degrassi, B.~A.~Kniehl and A.~Sirlin,
  Phys.\ Rev.\  D {\bf 48}, 3963 (1993).

\bibitem{singletcollideral}
  V.~Barger, P.~Langacker and G.~Shaughnessy,
  Phys.\ Rev.\  D {\bf 75} (2007) 055013
  [arXiv:hep-ph/0611239];\\  K.~Cheung, J.~Song and Q.~S.~Yan,
the
  arXiv:hep-ph/0703149.

\bibitem{hto4sm}
S.~Chang, P.~J.~Fox and N.~Weiner,
JHEP {\bf 0608}, 068 (2006) [arXiv:hep-ph/0511250];\\
U.~Ellwanger, J.~F.~Gunion and C.~Hugonie,
 JHEP {\bf 0507}, 041 (2005)
 [arXiv:hep-ph/0503203];\\
M.~Carena, T.~Han, G.~Huang and C.~E.~M.~Wagner
 in preparation

\bibitem{higgssearchinprep}
  V.~Barger, P.~Langacker, M.~McCaskey, M.~J.~Ramsey-Musolf and G.~Shaughnessy,
  arXiv:0706.4311 [hep-ph].

\end{thebibliography}
\end{document}